\documentclass[iop]{emulateapj}

\newcommand{\etal}{{et al.~}}

\shorttitle{SCUPOL Magnetic Field Analysis}
\shortauthors{Poidevin, F.; Falceta, G. C.; Kowal, G., De Gouveia Dal
  Pina, E. M., Magalh\~{a}es, A. M.} 

\begin{document}  

\message{ !name(scupol_v11.tex) !offset(-3) }

\title{MAGNETIC FIELD COMPONENTS ANALYSIS OF THE SCUPOL 850 $\mu$m POLARIZATION DATA CATALOG}
\shorttitle{}
\shortauthors{}


\author{Fr\'ed\'erick POIDEVIN}
\affil{University College London, Kathleen Lonsdale Building, Department of Physics \& Astronomy, Gower Place, London WC1E 6BT, United Kingdom}
\email{Poidevin@star.ucl.ac.uk}

\author{Diego Falceta-Gon\c{c}alves}
\affil{SUPA, School of Physics \& Astronomy, University of St Andrews, North Haugh, St Andrews, Fife KY16 9SS, UK}
\affil{Universidade de S\~{a}o Paulo,- Escola de Artes, Ci\^encias e Humanidades
Rua Arlindo B\'ettio, no. 1000 - Ermelino Matarazzo - S\~{a}o Paulo - SP 03828-000 , Brazil}
\email{dfalceta@usp.br}

\author{Grzegorz Kowal}
\affil{Universidade de S\~{a}o Paulo,- Escola de Artes, Ci\^encias e Humanidades
Rua Arlindo B\'ettio, no. 1000 - Ermelino Matarazzo - S\~{a}o Paulo - SP 03828-000 , Brazil}
\email{kowal@astro.iag.usp.br}

\author{Elisabete de Gouveia Dal Pino}
\affil{Universidade de S\~{a}o Paulo, Instituto de Astronomia, Geof\'isica e C\^ien\c{c}as Atmosf\'ericas, 
Rua do Mat\~{a}o 1226, Butant\~{a}, S\~{a}o Paulo, SP 05508-900, Brazil}
\email{dalpino@astro.iag.usp.br}

\author{Antonio-M\'{a}rio Magalh\~{a}es}
\affil{Universidade de S\~{a}o Paulo, Instituto de Astronomia, Geof\'isica e C\^ien\c{c}as Atmosf\'ericas, 
Rua do Mat\~{a}o 1226, Butant\~{a}, S\~{a}o Paulo, SP 05508-900, Brazil}
\email{mario@astro.iag.usp.br}

\begin{abstract}

We present an extensive analysis of the 850 $\mu$m polarization maps 
of the SCUPOL Catalog produced by \citet{mat09}, focusing exclusively on the molecular clouds and star-forming regions.
For the sufficiently sampled regions, 
we characterize the depolarization properties and the turbulent-to-mean magnetic field ratio of each region.
Similar sets of parameters are calculated from 2D synthetic maps of dust emission polarization 
produced with 3D MHD numerical simulations scaled to the S106, OMC-2/3, W49 and
DR21 molecular clouds polarization maps.
For these specific regions the turbulent MHD regimes retrieved from the simulations, as described
by the turbulent Alfv\'en and sonic Mach numbers, are
consistent within a factor 1 to 2 with the values of the same
turbulent regimes estimated from the analysis of Zeeman
measurements data provided by \citet{cru99}.
Constraints on the values of the inclination angle $\alpha$
of the mean magnetic field with
respect to the LOS are also given. The values obtained from the comparison of the 
simulations with the SCUPOL data are consistent with the estimates made by use of 
two different observational methods provided by other authors.
Our main conclusion is that simple ideal, isothermal and non-selfgraviting 
MHD simulations are sufficient to describe the large scale observed physical properties of the 
envelopes of this set of regions. 

\end{abstract}

\newpage

\section{INTRODUCTION}
 
Many efforts have increasingly been made over the last century to describe and characterize the nature of the 
Insterstellar Medium (ISM) of our Galaxy. On the theoretical side some concepts proposed by \citet{kol41} 
have been of primary importance by providing a useful mathematical framework from which the ISM has firstly 
been described as an ideal magneto-hydrodynamic (MHD) turbulent fluid. From the observational point of view, 
the wide span of temperatures and densities has been divided in various ranges designated as components 
or phases of the ISM \citep[e.g.][]{cox05} within which various structures like large-scale structures 
of bubble walls, sheets and filaments of warm gas, subsheets and filaments of cold dense material, have 
been classified. Following the evolution of the observational technics and the increasing amount of data  
they have been providing, various models and more recently MHD numerical simulations have been explored 
in an attempt to explain and predict the dynamic evolution of the ISM and the formation of Giant Molecular 
Clouds (GMCs). 

The formation and evolution of GMCs is still a subject of strong debate. One of the main issues is to unveil the 
conditions that will lead to formation of cores; the cradles where star formation takes place.
Two classes of models for explaining GMC formation have been proposed. The top-down models investigate   
the formation of GMCs as triggered by large-scale gravitational, thermal and magnetic instabilities in 
the differential rotating disk of a galaxy  \citep[e.g.][]{kim02}. On smaller scales, the bottom-up models 
explore formation of GMCs by compression of substructures of the ISM by supernova remnants, shocks produced 
by superbubbles or compression in converging flows in the ISM
\citep[e.g.][]{hei09,van07,vaz11}. These
shocks can ultimately trigger star formation in these regions \citep[e.g.][]{mel06,lea09}.

Molecular clouds and star-forming regions and the physical
characterization of the finite structures and sub-structures of GMCs
are the main purpose of this work. They are part of the dense cold gas phase of the 
ISM characterized by densities above about 10 cm$^{-3}$ and temperatures below 100 K. 
While the amount of dust grains pervading such regions is only about 1$\%$ of the gas mass, 
their polarized thermal emission observed at submillimeter (submm) wavelengths provides crucial 
information regarding the magnetic fields. Based on current advancements it is believed that some of the 
dust grains are elongated and have a specific orientation with respect to the local magnetic field they 
are pervading, therefore submm polarimetry gives us information about the average magnetic field along the 
observed Line-Of-Sights (LOSs). \citet{laz07} gives an interesting review about the advancements of dust 
grain alignment theory \citep[see also][]{hoa12,and12}.
  
\citet{cha53} referred to visible polarimetry and interpreted the large scale dispersion of the magnetic 
field observed in the Galactic plane as fluctuations of the magnetic field lines departing from a well 
ordered Galactic plane uniform component. Based on MHD arguments they established a relation where the 
velocity of the transverse velocity wave is proportional to the intensity of the magnetic field and 
inversely proportional to the square root of the density of the medium, leading to estimates of the 
field strength of order of 1-10 $\mu$G. This Chandrasekhar and Fermi (CF) method became popular and has 
been lately transposed to smaller spatial scales in clouds envelopes and cores where submm 
polarimetry has made possible to probe the mean magnetic field orientation in structures 5 orders of 
magnitude denser than the difuse ISM. Many analyses lead to estimates
of the average plane of the sky (POS) component magnetic 
field strengths 2-3 order of magnitude higher than in the difuse ISM \citep[e.g.][]{gon90, hil09}. 
More recently the CF method with MHD simulations has been investigated and correction factors 
to the CF equation have been proposed \citep[e.g.][]{ost01,fal08}.

In addition to polarimetry, spectroscopy has been providing valuable information for characterizing
the magneto-turbulent properties of some clouds from the point of view of the gas; generally in their 
densest regions which allow for sensitive detections. 
Estimates of magnetic field intensities along some LOS have been successfully obtained by Zeeman 
effect measurements in various regions \citep[e.g.][]{cru99,hei09}. 
The Goldreich-Kylafis effect \citep[][]{gol81,gol82}
has also been successfully measured which shows that CO isotopes can also be polarized with
magnetic field orientations consistent with the ones inferred from polarized emission by 
dust grains at the scale of some cores \citep[e.g.][]{gir06,for08} or at galactic scales \citep[][]{li11}.
Using another complete different approach, emission spectroscopy of ions and neutrals from molecular clouds 
has been compared and analyzed \citep[][]{hou00, hou04b}. Based on reasonable assumptions, such analysis 
and further developements make possible to calculate the turbulent ambipolar scale in some regions 
\citep[][]{li08, hez10} which give important piece of evidence to theoretical arguments \citep[][]{mes56,str66}.
Such studies also provide important constraints on further modeling to explain how magnetic fields 
and turbulence combine with each other to slow down gravitational
collapse in molecular clouds \citep[see][]{san10,lea13}.
  
In this work we propose a new method for characterizing the magneto-turbulent properties of the envelopes 
of some Galactic molecular clouds;  by envelopes we mean the
sub-structures of the molecular clouds that surround the embedded
cores. This method is based on the comparison of parameters extracted 
from the analysis of observed submm polarization maps (section
\ref{data}) with a similar set of parameters extracted 
from simulated maps (section \ref{simulations}). The results obtained with our method are discussed 
and compared with other published analyses (section \ref{discussion}). 
 The data are from the SCUPOL catalog provided by \citet{mat09}.
The 1024$^{3}$ synthetic cubes obtained from three-dimensional (3D) MHD
simulations of the turbulent ISM which were 
used for making the maps follow the description given by \citet{fal08}. 
More discussion on the CF method and on Zeeman Splitting measurements are given in sections \ref{cf_vs_adf}
and \ref{zeeman}, respectively. A summary of our results and our
conclusions are given in section \ref{summary}.

\section{DATA ANALYSIS} \label{data}

The data discussed in this work come from the SCUPOL catalog produced by \citet{mat09}.
This catalog is the product of the analysis of all regions observed between 1997 and 2005 at 850 $\mu$m 
in the mapping mode with SCUPOL, the polarimeter for SCUBA on the James Clerk Maxwell Telescope (JCMT). 
All imaging polarimetry made in the standard "jiggle-map" mode was systematically re-reduced and among 104 regions, 
83 regions presenting significant polarization with a signal-to-noise ratio such that, $p/\sigma_{p} >2$, 
are compiled (where $p$ is the polarization degree and $\sigma_{p}$ is
the uncertainty on $p$). 
The various fields cover 1 region in the Galactic Center, 48 Star Forming Regions (SFRs), 11 Young Stellar 
Objects (YSOs), 6 Starless Prestellar Cores (SPCs), 9 Bok Globules, 2 Post AGB stars, 2 Planetary Nebula, 2 Supernova Remnants
and 2 Galaxies. 

\subsection{Selected Regions} \label{selreg}

All maps with a sample of detection lower than 30 pixels are systematically considered too small to be 
statistically significant and are not included in our analysis. This implies that all regions classified as Bok Globules, 
post AGB stars and Planetary Nebula are not included. 
Since our work is mainly focused on star-forming and molecular cloud regions the targets of the catalog classified as
Supernova Remnants and Galaxies are also not included into our analysis but the highly sampled Galactic Center region 
is for comparison with SFRs, YSOs and SPCs regions.
Some of the SFRs, YSOs and SPCs regions were rejected when the sample or the spatial distribution of the pixels 
was not allowing to make a proper second order structure function analysis of the polarization map (see section \ref{tad}). 
The selected SCUPOL catalog regions are displayed in column 1 of Table \ref{table_regions}.
The majority of the regions are classified as SFRs, as indicated in column 2 of the table. 
The number of pixels of each maps is displayed in column 3 in Table \ref{table_regions}.  
Distances provided by \citet{mat09} and references therein are displayed in column 4. 
In the case of OMC-1, the group of vectors centered around R.A.(J2000)$=5:35:30$ and
Dec(J2000)$=-5:20$ \citep[see Figure 25 in][]{mat09} were not included in the
analysis in order to allow direct comparisons with former analysis in the OMC-1 region \citep[e.g.][]{hil09}. 

\subsection{Inferred Parameters} \label{infpar}

In this section we introduce the various parameters inferred from the analysis of the polarization maps; 
i.e. the SCUPOL catalog's I, Q and U Stokes maps and the uncertainty maps provided by \citet{mat09}.
The parameters are used to characterize each observed region.
They are useful to make statistics on several type of regions.
They will be compared to similar sets of parameters extracted from the analysis of scaled simulated maps. 

For any sample of data $d$ obtained on a region we define $<d>$ as the mean value of the distribution and $s(d)$
as its dispersion around the mean value. For any distribution of inferred parameters $ip$ obtained 
from a set of maps, we define $\overline{ip}$, as the average value obtained over different maps.

\subsubsection{Mean Polarization Degree and Polarization Angle Dispersion} \label{pol_obs}

Dealing with linear polarization maps one commonly defines the means and the dispersions of the polarization degree
and of the polarization position angle distributions. Since the polarization position angle is a variable 
that wraps over itself the averages retained in our analysis correspond to the means obtained where the 
dispersions of the distributions are found to be the smallest. 
This method of calculation helps one to avoid to make any assumption about the combination of a simple or multiple Gaussian 
distribution that would characterize the large scale uniform magnetic field component and a random distribution 
that would characterize the turbulent magnetic field component \citep[e.g.][]{goo90}. 

The definitions of the polarization degree, $p$, and of the polarization angle $\theta$ and the uncertainty 
maps of $\sigma_{p}$ and $\sigma_{\theta}$ follow equations (1)-(5) of \citet{mat09}. 
The values of $<p> \pm$ $ s(p)$ and $<\theta_{p}> \pm$ $s(\theta_{p})$ calculated for the 
regions retained in our analysis are displayed in columns 5 and 6 in Table \ref{table_regions}, 
respectively. The mean polarization position angles are given in the Galactic frame
and are positively counted from North to East. 

\begin{figure}
\epsscale{1.}
\plotone{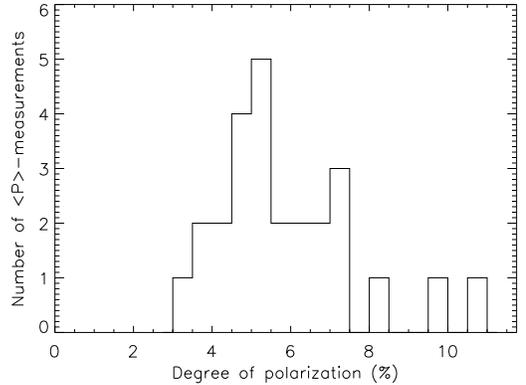}
\caption{Histogram of the mean polarization degree, $<p>$, of the sample including SFRs, YSOs and SPCs.
\label{SFR_pmeanhisto}}
\end{figure}

\begin{figure}
\epsscale{1.}
\plotone{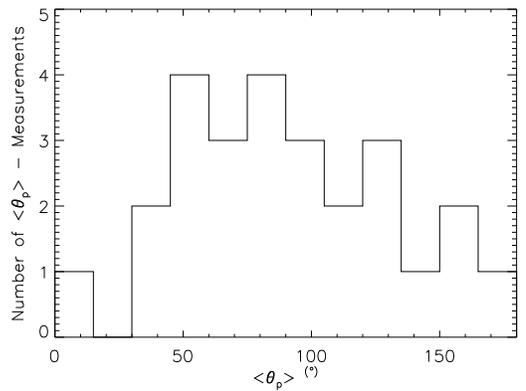}
\caption{Histogram of the mean Galactic polarization position angles, $<\theta_{p}>$, of the sample including SFRs, YSOs and SPCs.
\label{SFR_thetameanhisto}}
\end{figure}

The histogram of the mean polarization degree of the data set including the SFRs, the YSOs and the SPCs 
is shown in Figure \ref{SFR_pmeanhisto}. It shows values of $<p>$ lying between $3\%$ and $11\%$. 
The histogram of the mean Galactic polarization position angles of the same set of data is shown 
in Figure \ref{SFR_thetameanhisto}. It shows an
avoidance of low PAs and suggests a broad peak, but given the small size of
our sample (n=27 objects) 
it could also be indicative of no specific orientation of the mean magnetic field
orientations with respect to the Galactic plane. 
Such a comparison is actually out of the scope of the present work,
but we find that our result could be consistent with the 
conclusions of the detailed analysis conducted by \citet{ste11} on a sample of 52 Galactic star forming regions
observed at 350 $\mu$m. 

\subsubsection{Variations of Polarization with Intensity: The Depolarization Parameter} \label{pvsi}

The variations of the polarization degree, $p$, with the flux density, $I$, of dust grains emission  
is generally described by a power-law relation of the form $p \propto I^{\gamma}$ \citep[e.g.][]{gon05, mwf01}. 
In general the power index, $\gamma$, is negative which translates as a decrease of the polarization
with an increase of the intensity. This parameter is well suited to characterize the well-known ``polarization 
hole'' problem frequently observed along filaments with embedded cores
\citep[e.g.][]{dot96, hil99}.
For this reason, whatever the value it will take (positive or negative), in the following we refer to this parameter 
as the depolarization parameter. At the scale of a molecular cloud $\gamma$ takes different values 
according to whether the analysis considers the envelopes or the cores \citep[][]{poi10}. 
Initially, we systematically make estimates of $\gamma$ for each map
of the regions in the sample.
The values are displayed in column 7 of Table \ref{table_regions}. We find strong variations of $\gamma$. 
The lowest value is $\gamma = -1.44$ in the YSO L43 and the highest value, a 
positive one, is $\gamma = 0.10$ in the SFR OMC-1. More discussion on the variation of this 
parameter with column density variations is provided in section \ref{simstats}.

\subsubsection{Turbulent Angular Dispersion Parameter}  \label{tad}

\begin{figure*}
\centering
\includegraphics[width=8cm]{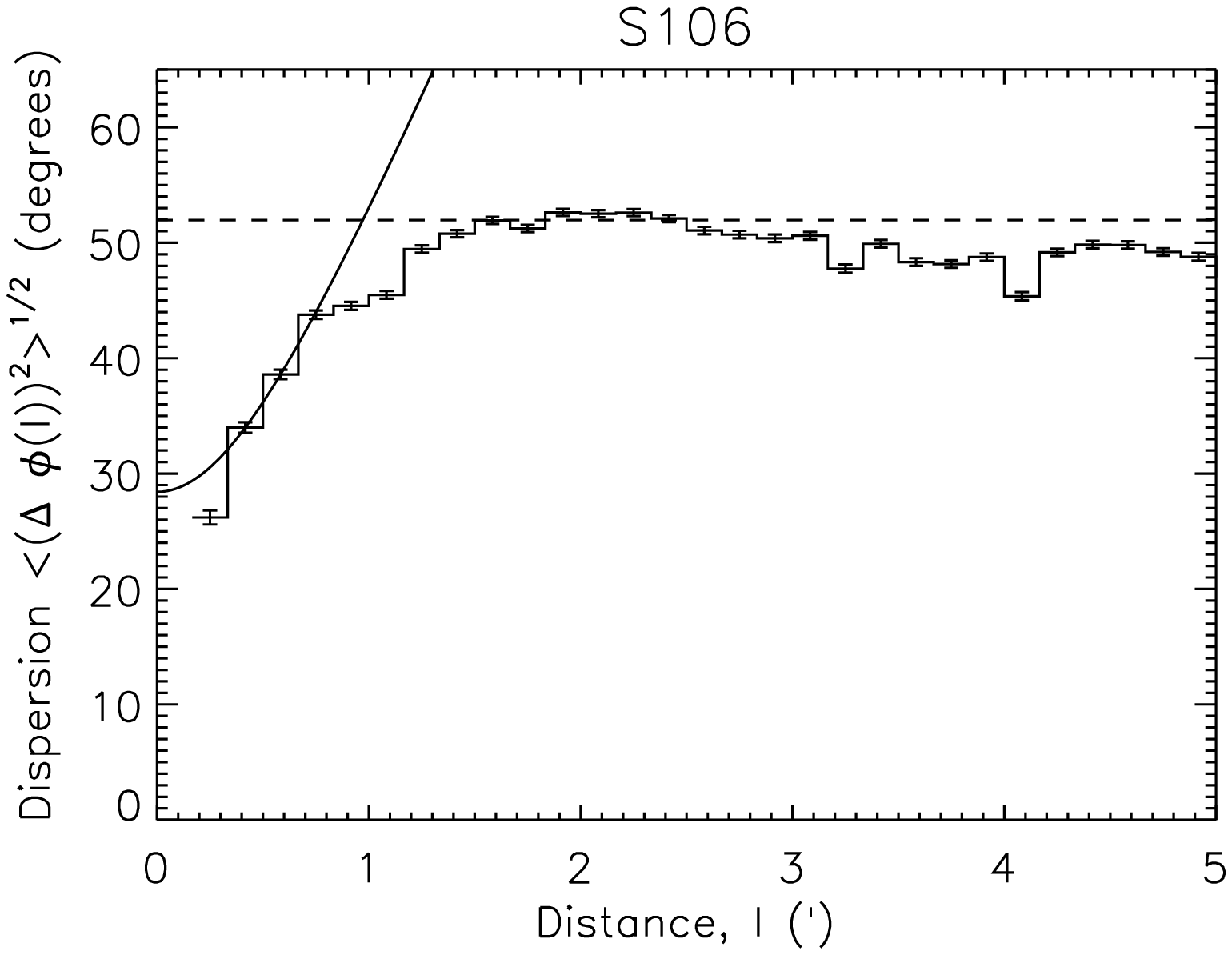}
\includegraphics[width=8cm]{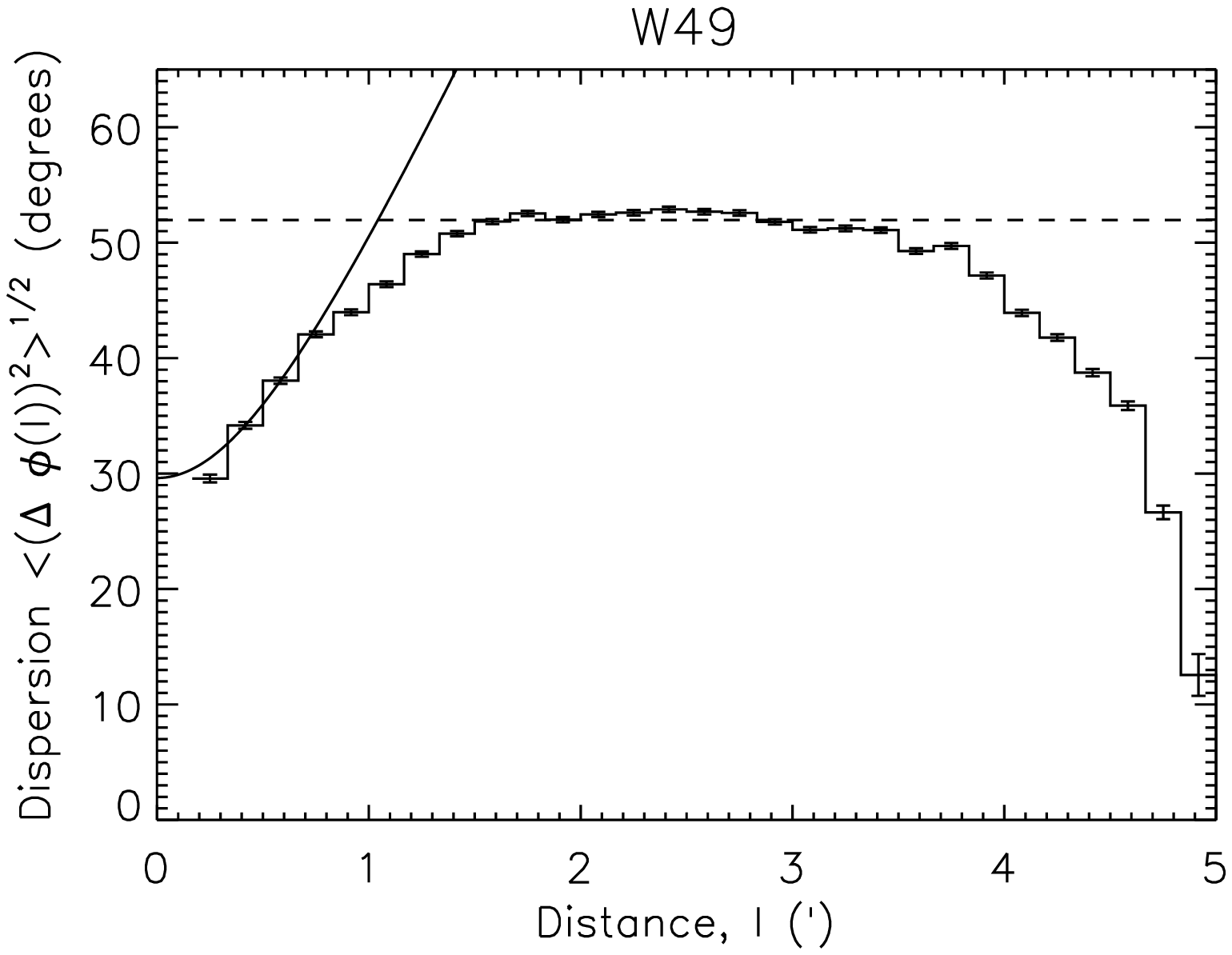}\\[10pt]
\includegraphics[width=8cm]{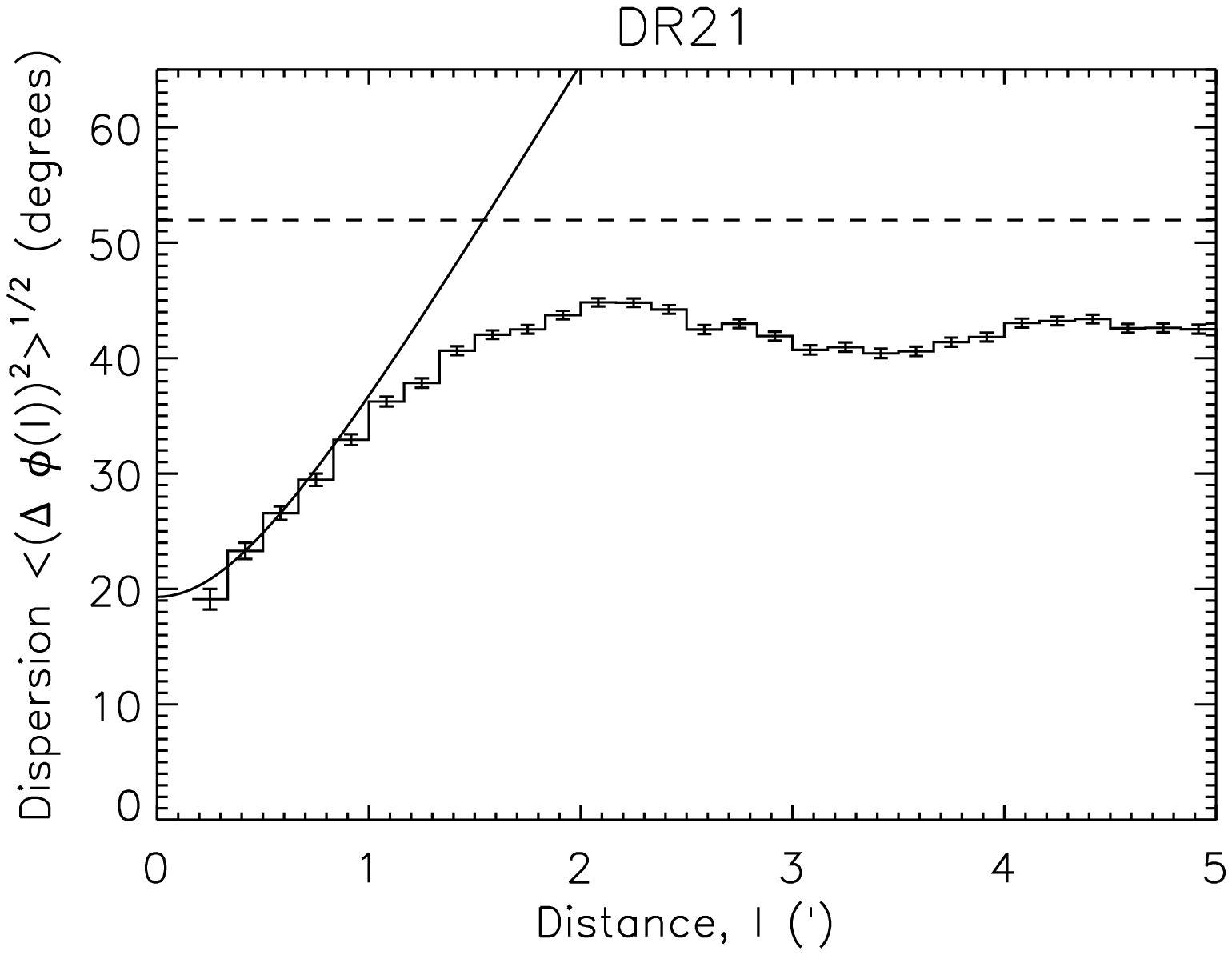}
\includegraphics[width=8cm]{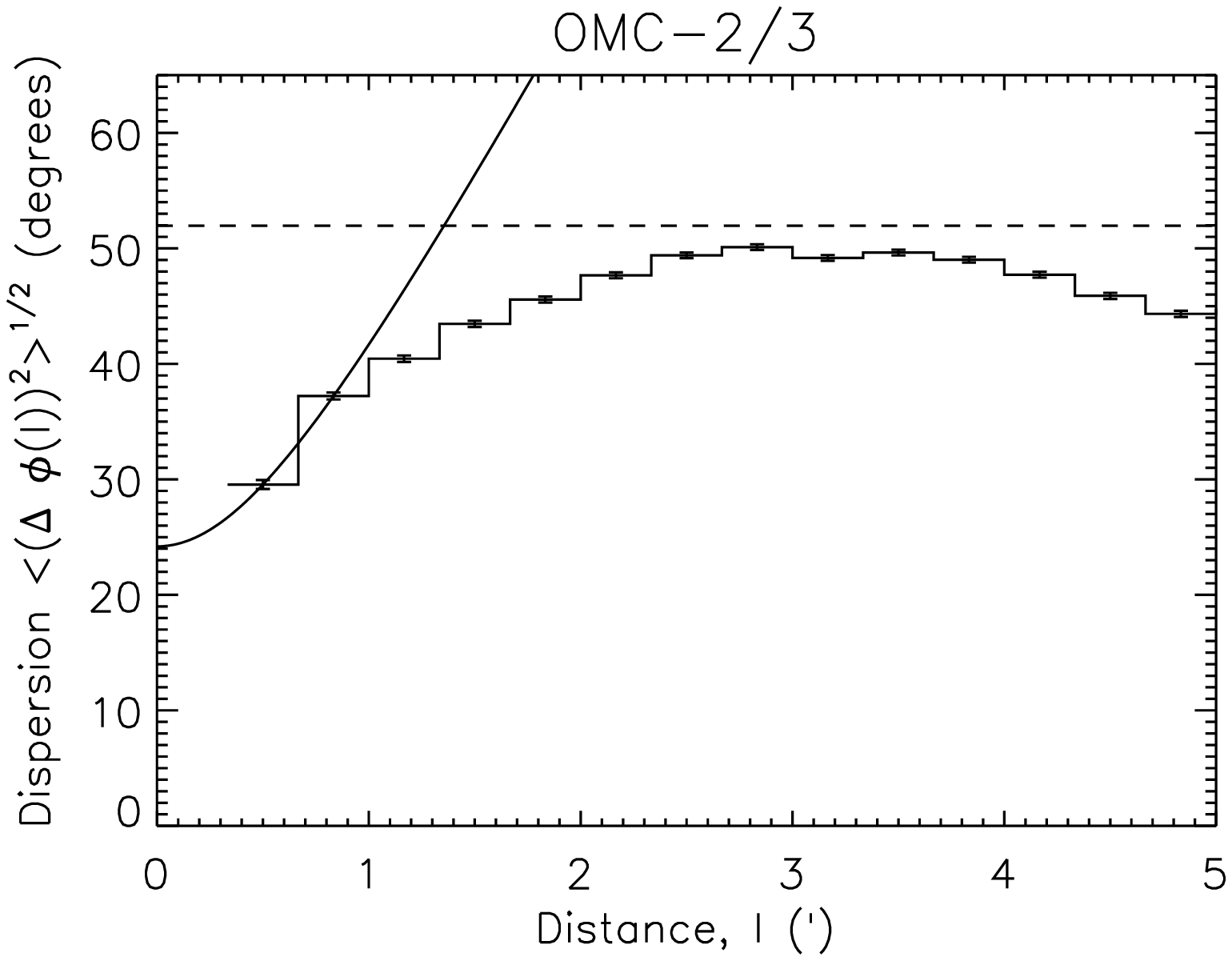}
\caption{Angular dispersion function, $<\Delta \theta^{2}(l)>^{1/2}$, for
  S106, W49, DR21 and OMC-2/3. The turbulent contribution to the total
  angular dispersion is determined by the zero intercept of the fit to
  the data at $l$=0.
}
\label{plotsf}
\end{figure*}

The Second-Order Structure Function (SF) of the polarization angles
obtained with measurements in the far-infrared - submm domain 
was first introduced by \citet{dot96}. The SF
gives the measurements of the autocorrelation of the polarization position angles 
$<\Delta \theta^{2}(l)>$ as a function of the distance $l$ measured for all pair 
of points into a map.  The square-root of the SF, also called the Angular Dispersion Function (ADF),  
can be used for determining the dispersion of magnetic field vectors about large-scale fields in
turbulent molecular clouds as it has been firstly proposed by \citet{fal08}, theoretically, and \citet{hil09} and \citet{hou09} with applications 
on the regions OMC-1, M17 and DR21. For applications on other regions see, for example, \citet{fra10} and \citet{poi10}.
Here we systematically use this method on the sample of regions shown in Table \ref{table_regions}.
The values of the turbulent angular dispersion parameter, $b$, which is the total angular dispersion 
determined by the intercept of the fit to the ADF at $l=0$ is shown in
column 8 of the table.
Examples of the fitting are shown with the plots in Figure
\ref{plotsf} for regions S106, W49, DR21 and OMC-2/3.
For these regions, the physical scales sampled are of size 
of about 29, 553, 145 and 40 mpc, respectively.
The effective beam size being of $22.5^{\arcsec}$, 
the fits have to be obtained on points with values of $<\Delta \theta^{2}(l)>$ 
estimated at $l$ about equal or greater than the effective beam size
\citep[see][]{hou09}. Therefore, for the maps which pixels are of size 
$20^{\arcsec} \times 20^{\arcsec}$ (e.g. OMC-2/3) the models were fitted 
on the first two point of the plots.
For most of the other maps which pixels are of size $10^{\arcsec} \times
10^{\arcsec}$ (e.g. S106, W49 and DR21)
the model was fitted on the estimates of the angular dispersions
obtained at $l=20^{\arcsec}$ and $l=30^{\arcsec}$. 
Having parameter $b$ we used Equation (7) of \citet{hil09} to estimate the 
ratio of the turbulent to the large scale magnetic field,
$\frac{<B_{t_{||}}^{2}>^{1/2}}{B_{0}}$, which is given
in column 9 of Table \ref{table_regions}. We find estimates of $b$ lying between $14.4^{\circ}$ in 
NGC 2024 and $44.5^{\circ}$ in Mon IRAS 12 implying turbulent to large scale magnetic field ratios 
lying between $18\%$ and $66\%$, respectively.
 
We point out that the values of $\frac{<B^{2}_{t_{||}}>^{1/2}}{B_{0}}$ shown in the last column of Table \ref{table_regions}
do not take into account the effect of the signal integration through the thickness of the cloud across
the area subtended by the telescope beam. Correction of such effects has been proposed by \citet{hou09},
but we are not using their method at this stage of our analysis. In the following, we adopt a 
complementary point of view, and rather than trying to remove the effects mentioned above,
we directly compare the sets of parameters ($<p>$, $s(\theta)$, $\gamma$ and $b$) extracted from 
the observed maps to similar sets of parameters inferred from synthetic maps that have been obtained 
from 3D MHD simulations scaled to the observations. This analysis is presented in section \ref{simulations}.


\subsection{Statistical Results} \label{obsstats}

\subsubsection{Parameter Variations with Distance and Size Sample}  \label{disteffects}

\begin{figure}
\epsscale{1.}
\plotone{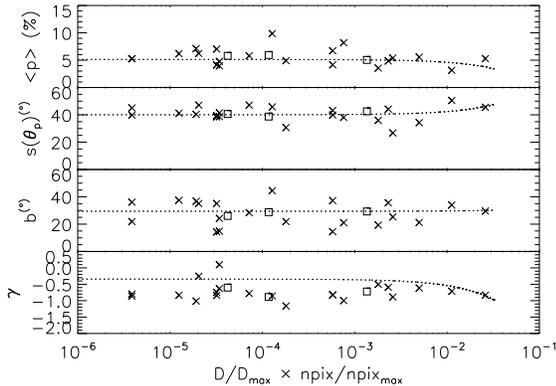}
\caption{Distribution of parameters $<p>$, $s(\theta)$, $\gamma$ and $b$ obtained for the subset of SFRs 
as a function of the normalized distance to each region combined with the normalized number of pixels of 
each observed map. Dotted-lines show linear fits to the distributions.
As it is suggested by the fits, the distribution of the data after binning in cells of size N=7 elements  
(square symbols) show very smooth variations for increasing values of $D/D_{\rm max} \times Npix/Npix_{\rm max}$.
\label{par_vs_dist_npix}}
\end{figure}

To study possible effects due to the combination of distance with map coverage, 
we define the normalized distance multiplied by the normalized number of pixels of our sample, 
$D/D_{\rm max} \times Npix/Npix_{\rm max}$.
The variation of parameters $<p>$, $s(\theta_{p})$, $b$ and $\gamma$ obtained for the SFRs sample (N=21) is plotted 
as a function of $D/D_{\rm max} \times Npix/Npix_{\rm max}$ in Figure \ref{par_vs_dist_npix}.
Linear fits to the distributions are plotted with dashed-lines showing very smooth variations of each parameter for values of 
$D/D_{\rm max} \times Npix/Npix_{\rm max} \le 2 \times 10^{-3}$. Also shown with square symbols
are the averages of the data in bins of size equal 7. Once binned the data also show very smooth variations
for increasing values of $D/D_{\rm max} \times Npix/Npix_{\rm max}$.

Our interpretation of these results is that the sample of SFRs is, at first approximation, quite homogeneous 
with respect to the distance to each source combined with the number of pixels in each map, therefore
the combined effects of the distance and of the map pixels sample size should introduce only a negligible 
statistical bias in our analysis, if any. We rather expect that the variations of the parameters obtained 
from one region to the other are primarily based on physical effects taking place in each region. 

\subsubsection{Dependence of the Parameters with Respect to Each Other}  \label{parspace}

The averaged values of the parameters discussed in the preceeding section 
are displayed in Table \ref{table_statistics_regions} for several subset regions. 
As mentionned previously, the Galactic Center is considered as a subset itself.
The subset of YSO regions (N$=4$ regions) and the subset of SPCs (N$=1$) regions 
are obviously too small to be statistically significant, but they are used for comparisons 
with the larger subset of SFR regions (N$=21$) and in the following of this section we will 
focus mainly on this subset. 

\begin{figure}
\epsscale{1.}
\plotone{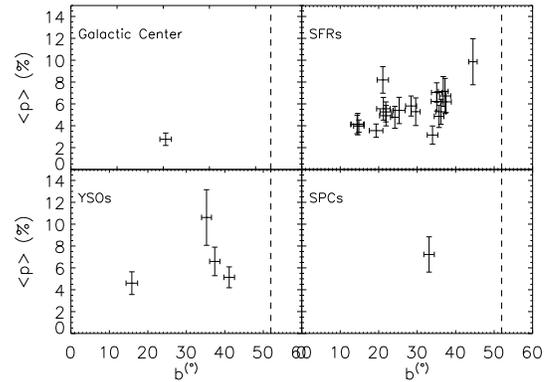}
\caption{Variations of $<p>$ with $b$ for the 4 subset regions. Values are displayed in Table \ref{table_regions}.
\label{pmean_vs_b}}
\end{figure}

\begin{figure}
\epsscale{1.}
\plotone{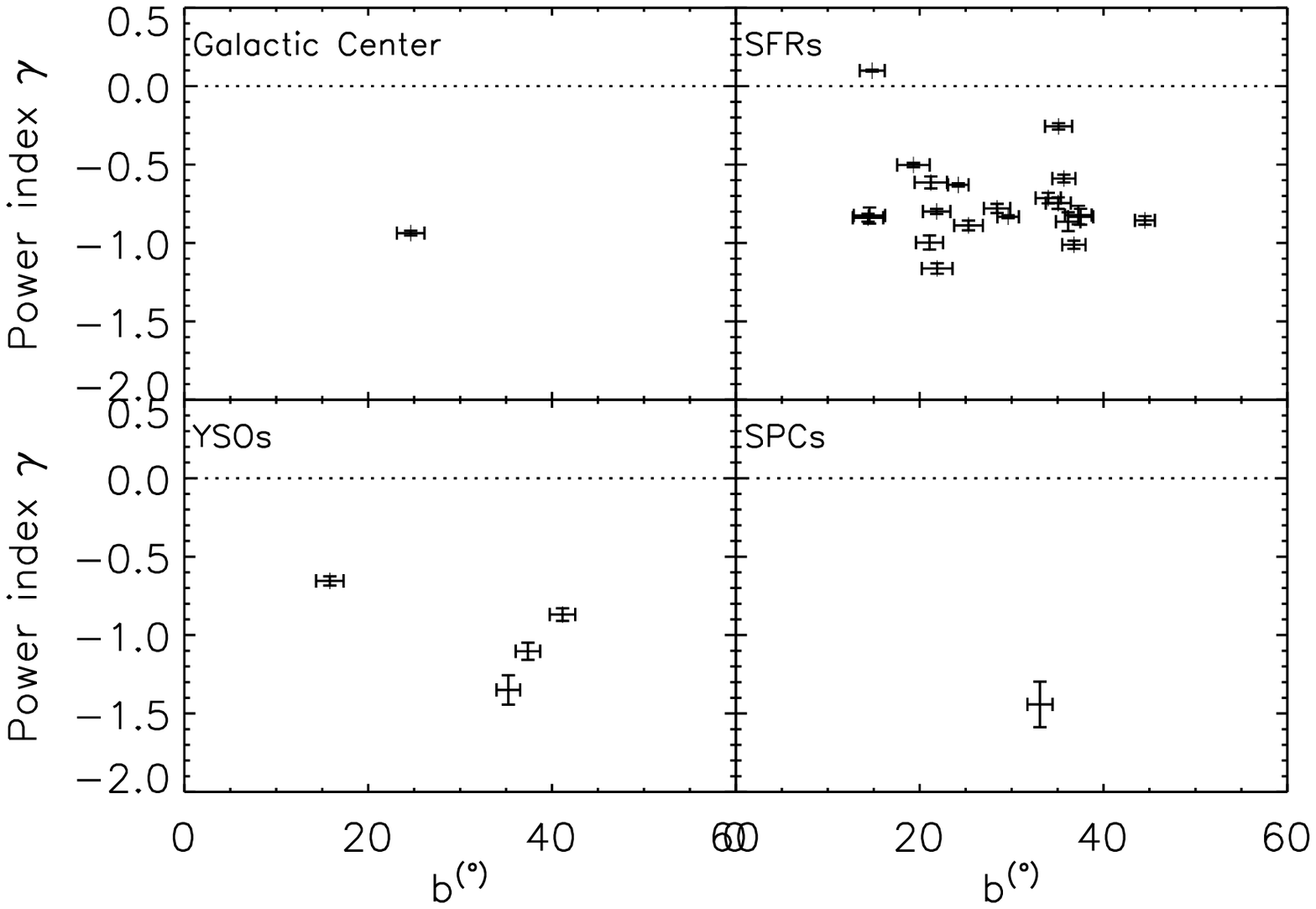}
\caption{Variations of $\gamma$ with $b$ for the 4 subset regions. Values are displayed in Table \ref{table_regions}.
\label{gamma_vs_b}}
\end{figure}

\begin{figure}
\epsscale{1.}
\plotone{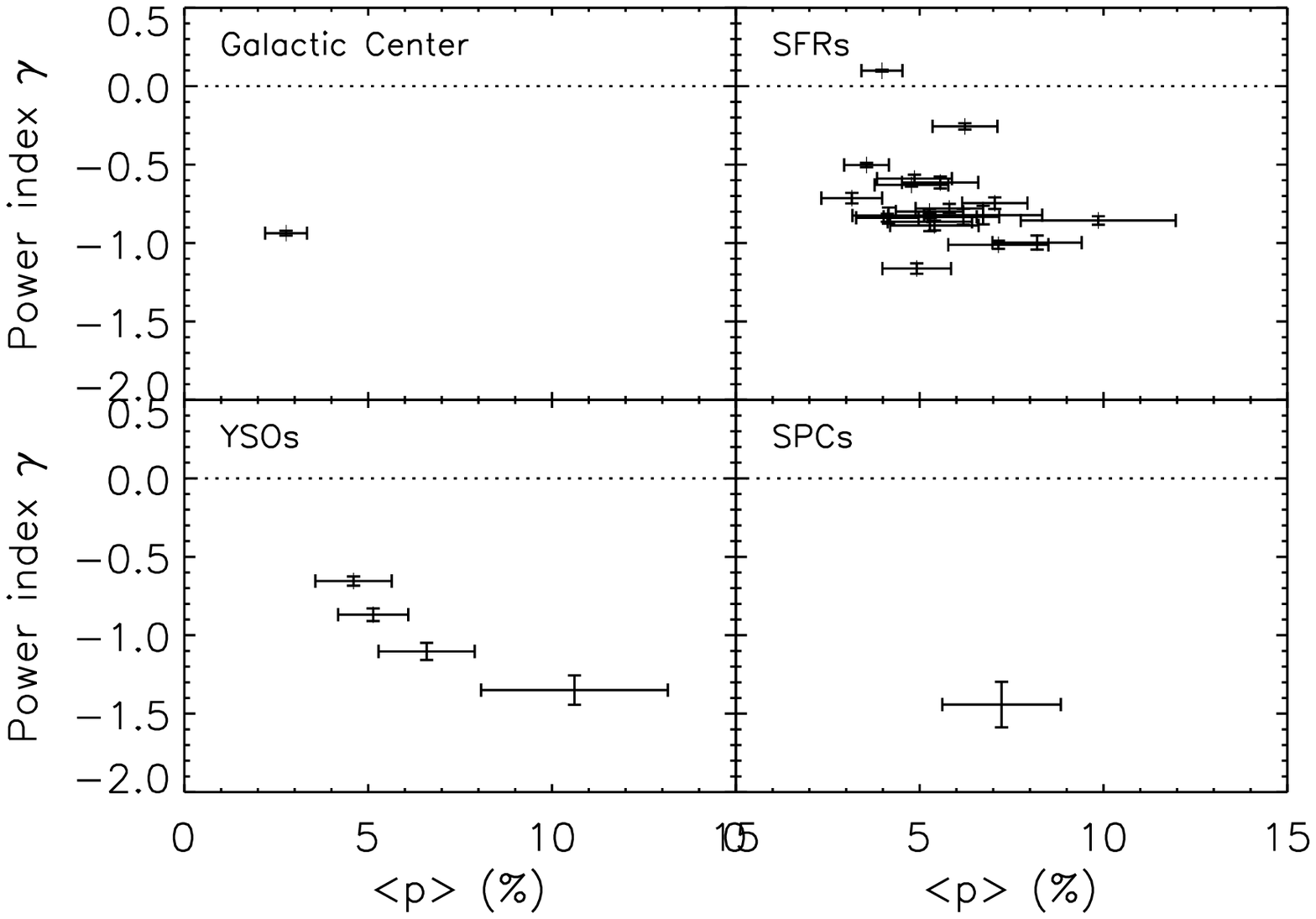}
\caption{Variations of $\gamma$ with $<p>$ for the 4 subset regions. Values are displayed in Table \ref{table_regions}.
\label{gamma_vs_pmean}}
\end{figure}

The variations of $<p>$ with $b$ are shown in Figure \ref{pmean_vs_b} for the 4 subset 
regions. The subset of SFRs shows large variations of $<p>$ with $b$ centered on
the averaged values $\overline{<p>}=5.59 \% \pm 1.59 \%$ and $\overline{b}=28\fdg0 \pm 8\fdg9$.
The plots suggest a slight increase of $<p>$ with the increase of $b$, however, since the
statistical analysis conducted by \citet{ste11} suggests that the magnetic field in 
molecular clouds is decoupled from the large scale Galactic magnetic field, we interpret 
the trend observed in the upper right part of Figure \ref{pmean_vs_b} to be rather
statistical in nature. Similarly, we do not find any correlation between parameters 
$\gamma$ and $b$ which is shown in Figure \ref{gamma_vs_b}, and between parameters $\gamma$ 
and $<p>$ which is shown in Figures \ref{gamma_vs_pmean}. For the subset of SFRs, the distribution 
of $\gamma$ is centered on $\overline{\gamma}=-0.74$ with a dispersion of $0.27$.
 
\begin{figure}
\epsscale{1.}
\plotone{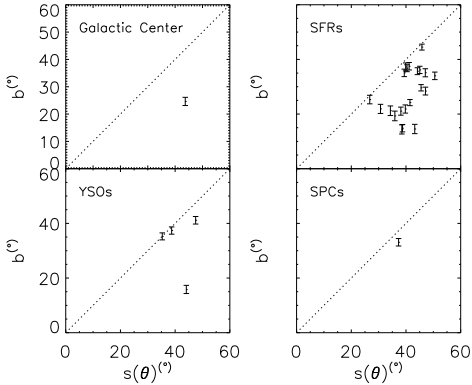}
\caption{Variations of $b$ with $s(\theta_{p})$ for the 4 subset
  regions. Values are displayed in Table \ref{table_regions}.
The dashed line corresponds to the values for which both parameters are equal.
\label{b_vs_stheta}}
\end{figure}

The variations of $b$ with $s(\theta_{p})$ are shown in Figure \ref{b_vs_stheta}.
The dashed line shows the location where the two parameters are equal. We find values of $b$ always 
lower or equal to those of $s(\theta_{p})$ and, as expected with the methods used to derive the two parameters, 
none of the regions show values of $b > s(\theta_{p})$. 
For the subset of SFRs the values of the turbulent angular dispersion $b$ 
are approximately three times smaller than the values of the dispersions around the global magnetic fields $s(\theta_{p})$.
According to whether one uses $b$ or $s(\theta_{p})$ such variations in the ratio of $b/s(\theta_{p})$ will 
introduce variations in the products of the Chandrasekhar \& Fermi method \citep[][]{cha53, ost01, hou04, fal08, hil09}. 
Since $b$ should be more accurate a parameter than $s(\theta_{p})$ to estimate the small scale angular 
dispersion of a region, in the following we will use $b$ for determining the POS angular dispersion more 
generally expressed by $\sigma (\theta_{||})$.
 
\subsection{The CF Method}  \label{cf_vs_adf}

\citet{cha53} defined a method for estimating the strength of the POS magnetic field component, $B_{\rm pos}$, based on the 
POS angular dispersion, $\sigma (\theta_{||})$ and the one-dimensional velocity dispersion, $\sigma (v_{\perp})$, 
of the gas of mass density, $\rho$. The turbulence of the medium is supposed to be isotropic and the gas to be coupled 
to the Alfv\'enic perturbations, so that equipartition between the
kinetic and the perturbed magnetic 
energies is required. 
The Alfv\'en speed is given by
\begin{equation} \label{va}       
V_{A}=\frac{B_{0_{||}}}{\sqrt{4 \pi \rho}},
\end{equation}
and, in the small angle limit, the ratio between the mean turbulent magnetic field component 
in the POS, $<B_{t_{||}}^{2}>^{1/2}$, and the uniform component of the magnetic 
field, $B_{0_{||}}$ ($= B_{\rm pos}$ in this case), 
is given by
\begin{equation} \label{cf1}       
\frac{<B_{t_{||}}^{2}>^{1/2}}{B_{0_{||}}} \simeq \frac{\sigma (v_{\perp})}{\sigma(\theta_{||})}.
\end{equation}

\citet{fal08} conducted 3D Magneto-Hydrodynamic (MHD) simulations 
of turbulent interstellar medium regions
to create synthetic 2D polarization maps.
Their results show that equipartition between magnetic and kinetic energies is a fulfilled assumption 
for sub-Alfv\'enic models (i.e., for models where the turbulent velocity is smaller than the Alfv\'en velocity)
as well as for super-Alfv\'enic models after a few crossing times. 
Following \citet{cha53} they proposed  
\begin{equation} \label{cf2}
B_{\rm pos}=B_{0_{||}}+<B_{t_{||}}^{2}>^{1/2},
\end{equation}
with
\begin{equation} \label{cf3}
B_{\rm pos} \simeq C \sqrt{4 \pi \rho} \frac{\sigma(v_{\perp})}{\tan(\sigma(\theta_{||}))},
\end{equation}
to take into account $<B_{t_{||}}^{2}>^{1/2}$ and to avoid the small angle approximation. 
$C$ is a correction factor proposed by \citet{ost01} based on their MHD simulations.
These authors conclude that $C \sim 0.5$ is deemed appropriate in most cases 
to estimate $B_{\rm pos}$ to the condition that the field is not too weak.  
Shortcomings of the CF method are discussed by \citet{hou04} and a few technical and physical 
reasons are given to explain this correction factor. More recently, \citet{hil09} used 
the second order SF to make a two dimensional analysis of polarization maps and derived 
the following expression 
\begin{equation} \label{cf4}
B_{\rm pos} \simeq \frac{\sqrt{2-b^2}}{b} \sqrt{4 \pi \rho} \sigma(v_{\perp}). 
\end{equation}

Combining equation \ref{cf1} with equation \ref{cf3} or equation \ref{cf4} gives the general relation
\begin{equation} \label{ratio}
\frac{<B_{t_{||}}^{2}>^{1/2}}{B_{\rm pos}} \propto W 
\end{equation}
where 
$W= W1 = C \frac{\tan(\sigma(\theta_{||}))}{1-\tan(\sigma(\theta_{||}))}$ (equation \ref{cf2}),
$W = W2 = C \frac{\sigma(\theta_{||})}{1-\sigma(\theta_{||})}$ (equation \ref{cf2} at small angle limit) 
or
$W = W3 = \frac{\sigma(\theta_{||})}{\sqrt{2-\sigma(\theta_{||})}}$ (equation \ref{cf4}).

\begin{figure}
\epsscale{1.}
\plotone{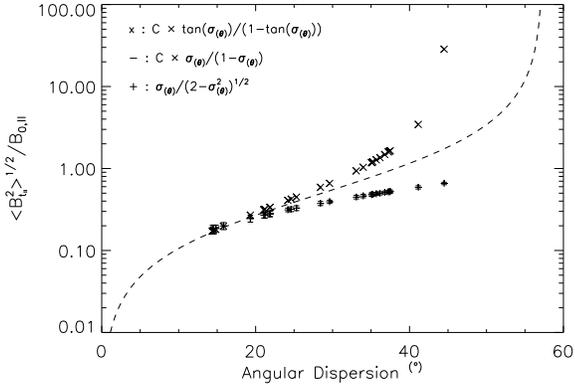}
\caption{Variations of the ratio of the POS mean turbulent magnetic field component to the POS mean uniform field component for the three cases described in section \ref{cf_vs_adf}. The dashed line gives the small angle limit ratio (see text for more details).
\label{ratios}}
\end{figure}

As a numerical check we used the values of $\sigma(\theta_{||}) = b$ displayed in Table \ref{table_regions}
to plot the variations of $W1$ and $W3$ as a function of  $\sigma(\theta_{||})$ in Figure \ref{ratios}.
The correction factor $C=0.5$ is used in the process and we find good agreement between the 2 
ratios for angular dispersions lying between $0^{\circ}$ and $20^{\circ}$. For comparisons the 
small angle limit ratio $W2$ is shown by the dashed-line. Discrepancies between the 3 ratios 
appear for angular dispersions higher than $\approx 20^{\circ}$. In this domain range, the CF method 
based on equation \ref{cf2} gives values higher than the small angle limit method while the 
CF method based on the SF approach and equation \ref{cf4} returns lower values than those obtained 
with the small angle limit method. The main reason for this discrepancy was pointed by \citet{fal08}. 
A key argument in their modeling is that the amplitude of the underlying reference magnetic 
field is also perturbed by the turbulent field, and the values of the uniform components are 
typically smaller than previously estimated by a factor that equals the turbulent component (see discussion below).

\subsection{Zeeman Splitting versus CF Method} \label{zeeman}

Zeeman splitting measurements are available for some of the regions of the SCUPOL catalog.
The data we refer to come from a survey conducted by \citet{cru99} including emission and absorption 
observations, as well as VLA synthesis observations. The analysis provides averaged LOS magnetic field 
strengths on the area sustended by the beams unless the targets are point source like. 

For the regions where polarimetry obtained with SCUBA and Zeeman measurements provided by \citet{cru99}
are both available, we have combined the densities and velocity dispersions values obtained from Table 1 of \citet{cru99}
with our estimates of $b$ shown in Table \ref{table_regions} and Equation \ref{cf4} was 
used for estimating averaged POS magnetic field intensity components over the area of the clouds.
A summary of the data used in the process is shown in Table \ref{table_bfields}.
The LOS magnetic field components estimated by Crutcher are displayed in column 1 and our  
estimates of the POS components are shown in column 4 (see values with
no parenthesis). 
We find a mean ratio of 4.7 between the POS and LOS magnetic field components obtained with the two methods.
The sample shows a dispersion to the mean of 2.8. 

\begin{figure}
\epsscale{1.}
\plotone{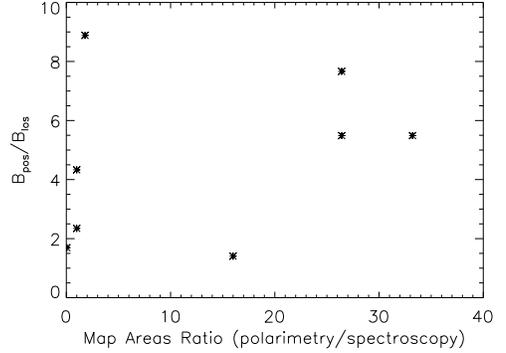}
\caption{Variations of the ratio of the POS to LOS magnetic field strengths as a function of the effective areas of the maps used to estimate the two magnetic field intensities, respectively.   
\label{bratiosmaps}}
\end{figure}

Several arguments could explain 
why our estimates of B$_{\rm POS}/$B$_{\rm LOS}$ are systematically greater than one. 
A first naive one could be that we are probing magnetic fields in a sample of clouds where the large 
scale uniform component is always closer to the POS ($\alpha \ge
55^\circ$) than to the LOS. One another possibility could be that the areas subtended to estimate
$B$ with each method, if too much different, would introduce a bias on
the estimates of B$_{\rm POS}/$B$_{\rm LOS}$. 
To look on this side, we made calculations of the ratio of the areas covered by polarimetry and by spectroscopy, respectively. 
Appendix A of \citet{cru99} and references therein were used to define the values of the map areas observed for
making Zeeman splitting estimates. For single-antenna observations of absorption lines toward
continuum sources that are smaller than the telescope beam, the effective resolution is the angular 
size of the source therefore, we give the ratio between the map areas a value of one for these sources.  
The values of the ratios between the areas of the polarimetry and spectroscopy maps are shown in column 
6 of Table \ref{table_bfields}. We find a mean ratio of 13.2 between the areas used to make POS and LOS 
magnetic field components estimates, respectively. The sample has a
large dispersion to the mean of 13.9. 
Figure \ref{bratiosmaps} shows the distribution of the magnetic fields ratio as a function of the map 
areas ratio. If a bias on B$_{\rm POS}/$B$_{\rm LOS}$ was introduced
by an increasing ratio between the areas observed with
polarimetry and with spectroscopy, one could expect a correlation between
the two ratios. This seems not to be the case but we also point out
how limited would be any conclusion with such a
small sample. 

One another argument to explain the high values of 
B$_{\rm POS}/$B$_{\rm LOS}$ is that the Zeeman measurements 
could be subject to magnetic field reversals 
toward the LOS \citep[e.g.][]{poi11,kir09} which is an effect to which
the CF method should be completely blinded to. 
In such a case the estimates of B$_{\rm LOS}$ should be considered as lower limits.
In addition, LOS magnetic field components, as well as the parameters
displayed in Table \ref{table_bfields} 
are subject to spatial averaging effects \citep{cru99} so that the same 
values of $n$ and $\sigma_{v}$ for estimating the POS and LOS sky magnetic field components likely 
introduced a bias in our analysis. As an example, if one refers to the high 
resolution estimates of the inclination angles with the LOS, $\alpha$, for DR21 obtained by \citet{kir09},
inclination angles are found to lie between $28.3^{\circ}$ and $77.1^{\circ}$, while combining our 
values of B$_{\rm LOS}$ and B$_{\rm POS}$ gives an inclination angle
of about $81.2^{\circ}$. 
Actually, it is not clear, whether or not, OH or CO isotopologues trace similar gas
densities as dust. In addition, optically thin transition of molecular
ion like, H$^{13}$CO$^{+}$(J=3-2), might be more appropriate for
determining velocity dispersions \citep[e.g.][]{hil09}.  
A survey of H$^{13}$CO$^{+}$(J=3-2) published data gives velocity
dispersion values of 1.85 km/s \citep[][]{hou00,hil09} for OMC-1, 
about 0.68 km/s \citep[][once data plotted in their Figure 5 are
averaged]{li10} for NGC 2024, about 0.58 km/s \citep[][]{pon09} for
Rho Oph A toward the core N6, and of 2.04 km/s \citep[][]{hou00} for
DR21 (OH1) and DR21 (OH2). Those values are of order one to three
times higher than the ones obtained from CO or OH data, which means magnetic field
intensities B$_{\rm POS}$ of order one to three times higher than the
ones obtained from CO or OH when densities similar to the one displayed with
no parenthesis in 
Table \ref{table_bfields} are used for the calculations, or even
higher if one systematically considers densities $n>10^{5}$ cm$^{-3}$
(see data with parenthesis). 
On the other hand, except for NGC 2024, most of the
H$^{13}$CO$^{+}$(J=3-2) data have been obtained toward cores and might
be not representative of the averaged values one would get from
observations of this tracer over the
area sustended by the clouds, for that reason we consider both cases,
i.e. constraints on B$_{\rm POS}$ and therefore on $\alpha$, as
obtained from CO, OH or from H$^{13}$CO$^{+}$(J=3-2); see values given
with no and with parenthesis in Table \ref{table_bfields}, respectively. 
 
In addition, as was discussed by \citet{heil09}, the probability density functions used for comparing 
$B_{\rm LOS}/B_{\rm total}$ and B$_{\rm POS}/$B$_{\rm total}$
are such that half the time, B$_{\rm POS}/$B$_{\rm total} > 0.87$, which makes B$_{\rm POS}$ a better tracer of B$_{\rm total}$ than does B$_{\rm LOS}$. 
Following this line of thought, we could consider our estimates of
B$_{\rm POS}^{HIL09}$ as 
upper limits and we expect the estimates of B$_{\rm LOS}$ to be lower limits.  
We have displayed our estimates of $\alpha$ in the last column of Table \ref{table_bfields},
but considering the arguments above those should be considered only as upper limits.

Finally, it is possible that the POS magnetic field is systematically overestimated, depending on the 
turbulent regime. Therefore, we also present in Table \ref{table_bfields} values of 
B$_{\rm POS}^{FLK08}$, as estimated from Equation \ref{cf3}. Large differences between 
B$_{\rm POS}^{FLK08}$ and B$_{\rm POS}^{HIL09}$, based on \citet{hil09}, is likely to occur 
for dispersions $\Delta \theta > 30^\circ$. The values differ from a factor of 3 up to more than 
one order of magnitude. For example, the value obtained for OMC-1 is very high compared to the much smaller $\sim 760\mu$G obtained by Houde et al. (2009) using the same technique but by correcting for signal integration through the depth of the cloud and across the telescope beam - in better agreement 
with the estimate B$_{\rm POS}^{FLK08} = 395 - 1229 \mu$G. It is not clear though the actual 
significance of each method since the values of B$_{\rm POS}^{FLK08}$ for W49 and S106 seem too low and do not correspond well to the Alfv\'en turbulent component provided from the simulations, as explained in 
the next section. We must point that despite of the apparent overestimation of B$_{\rm POS}^{HIL09}$ 
in the cases of OMC-1 and DR21, these values, combined with B$_{\rm LOS}$, $n_{\rm H_2}$ and $\sigma_v$ 
shown in Table \ref{table_bfields}, are in good agreement with the alfv\'enic turbulence obtained 
from the numerical simulations.

\section{COMPARISON TO MHD NUMERICAL SIMULATIONS} \label{simulations}

\citet{fal08} modeled the statistics of polarized emission from 
dust grains based on MHD numerical simulations, and showed its strong dependence on the turbulent 
regime of the host molecular clouds. From an observational perspective then it is possible to 
estimate the turbulent regime of a given molecular cloud, as well as the orientation of the 
mean magnetic field with respect to the plane of sky, by comparing the statistics of the observed 
polarization maps to those obtained by numerical simulations with similar scaling.

\subsection{Scaling of the Simulations}

In order to estimate the magnetic field and turbulent regimes of part of the sample of the objects 
mentioned before, we performed a number of MHD numerical simulations of turbulence in molecular 
clouds, each of them related to a specific set of initial conditions chosen to best reproduce 
the observations of a given object.

The problem of magnetic turbulence in molecular clouds can be solved by a fluid approximation
governed by the isothermal ideal MHD equations of the form:

\begin{equation}
\frac{\partial \rho}{\partial t} + \mathbf{\nabla} \cdot (\rho{\bf v}) = 0,
\end{equation}

\begin{equation}
\frac{\partial \rho {\bf v}}{\partial t} + \mathbf{\nabla} \cdot \left[ \rho{\bf v v} + 
\left( p+\frac{B^2}{8 \pi} \right) {\bf I} - \frac{1}{4 \pi}{\bf B B} \right] = {\bf f},
\end{equation}

\begin{equation}
\frac{\partial \mathbf{B}}{\partial t} - \mathbf{\nabla \times (v \times B)} = 0,
\end{equation}

\begin{equation}
\mathbf{\nabla \cdot B} = 0,
\end{equation}

\begin{equation}
p = c_s^2 \rho,
\end{equation}

\noindent
where $\rho$, ${\bf v}$ and $p$ are the plasma density, velocity and pressure, 
respectively, ${\bf B = \nabla \times A}$ is the magnetic field, ${\bf A}$ is the 
vector potential and ${\bf f} = {\bf f_{\rm turb}}+{\bf 
f_{\rm visc}}$ represents the external source terms, responsible for 
the turbulence injection.

We solve the MHD equations above using a high order shock-capturing Godunov-type
scheme \citep[][]{kow09} based on a multi-state Harten-Lax-van Leer (HLLD) Riemann
solver for the isothermal MHD equations and a 4th order Runge-Kutta (RK) scheme for time
integration. The divergence of the magnetic field is kept close to zero by using a field
interpolated constraint transport (CT) scheme on a staggered grid, and periodic boundaries.
Turbulence is driven by a solenoidal and, therefore incompressible forcing, in Fourier space,
and random in time. The choice of the initial setup of the simulations depends on the 
observational data and is described below.

The turbulent regime of a given run is determined by the rms of the sonic and Alfvenic Mach 
numbers calculated for the entire cube. Statistically the regime is therefore, mostly 
related to the amplitude of the fluctuations at the injection scale. In the ISM, the injection 
of kinetic energy is believed to occur at scales larger than 10pc \citep[e.g.][]{arm95}.
The ISM turbulence presents a selfsimilar cascade over several decades on 
lengthscales, from injection down to sub-AU scales where dissipation occurs. The numerical 
simulations in a fixed grid of $1024^3$ cells present turbulent scales that goes from the largest
scales of energy injection in the ISM ($L = 10 - 50$pc) to the dissipation scales ($l_{\rm min} =
0.01 - 0.1$pc), where the last is basically related to numerical diffusivity or 
other physical mechanisms, e.g. on the ambipolar diffusion (see 
Falceta-Gon\c calves et al. 2010, Falceta-Gon\c calves \& Lazarian 2011). 
On the other hand, the observed turbulent scales of the molecular clouds lie in between the 
injection and the dissipation scales, i.e. few parsecs in length.
For obvious reasons, it is impossible to use one single simulation to compare all observational 
data and some sort of sampling of the numerical data is required. 

In order to properly select the model parameters the determination of the dynamical 
range of scales of a given observed cloud is crucial. For that we may make use of the 
assumption that the molecular cloud observed corresponds to part of the inertial range 
scales of a self-similar ISM turbulence. The turbulence at the largest
scales presents typical amplitudes of $\sim 10 c_{s,20}$ (Larson 1980), being $c_{s,20}$ the 
isothermal sound speed for $T = 20$K. Also, the Kolmogorov's model of turbulence gives the 
scaling of velocity amplitudes as $\left< \delta v^2_{l} \right>^{1/2} \propto l^{1/3}$. Therefore, combining the 
large scale amplitude with the given scaling it is possible to predict the turbulent amplitudes 
at any given lengthscale $l$. For instance, the region S106 is $\sim 1$pc wide, therefore the 
turbulent amplitude at its largest scale will be approximately given by $\left< \delta v^2_{l=2{\rm pc} } \right>^{1/2} \simeq 2 c_{s,20}$.

Regarding the numerical resolution used in comparing the cubes to the observed data one needs to 
determine the dynamical range of the turbulent motions given by $\mathcal{R} = l_{\rm cloud}/l_{\rm cut}$, where 
$l_{\rm cut}$ represents the largest of either the dissipation or the spatially resolved lengthscales. 
The dynamical range should be the same for the numerical and observational 
data (see Falceta-Gon\c calves et al. 2008). As an example, S106 presents $l_{\rm cut} \sim 0.01$pc, 
which gives $\mathcal{R} \sim 100$. Since the numerical diffusion causes the damping of the turbulence at scales 
of $\sim 10$ cells, it is only possible to simulate the observed polarization maps if the numerical 
resolution is of $1024^3$ cells. Combining the determined turbulence amplitude at the given scale of S106, 
and its dynamical range, we use a numerical setup with Mach number equals 2 and a numerical grid 
of $1024^3$ resolution.  

The same analysis has been repeated to OMC2-3, W49 and DR21, resulting in a total of 4 different setups 
for the numerical simulations (Table \ref{table_models_0p30}). For the
initial magnetic field we assume a uniform field, whose amplitude is equal to the equipartition value
at the largest scale, i.e.  $B_0^2 \sim <\delta v^2_{l=L}>$.

Interestingly, the values of sonic and Alfvenic Mach numbers presented in Table
\ref{table_models_0p30}, obtained assuming a self-similar turbulent cascade
from the large ISM scales down to the scales of the OMC-2/3, DR21, S106 and W49
regions, are in good agreement, within a factor of two, with the
observational estimates made by \citet{cru99} based on Zeeman measurements.

\subsection{Statistics of Observations vs Numerical Simulations} \label{simstats}

In order to compare the statistics of the observed regions with the numerical simulations we have to calculate the synthetic polarization maps for these data. Here we use an approximate radiative transfer model, in which we assume that the radiation is originated exclusively by thermal emission from dust grains and the medium to be optically thin to this radiation. The dust abundance is supposed to be linearly proportional to the gas 
density, and the dust particles distribution to be isothermal. 

Naturally the alignment of prolate and oblate dust particles with
respect to the magnetic field lines is not perfect in molecular
clouds. The physical mechanisms of grain alignment are not in the
scope of this study, so we assume a constant polarization efficiency,
$\epsilon = 0.1$, for starting. The local, i.e. in each cell, angle of alignment ($\psi$) is determined by the local magnetic field projected into the plane of sky, and the linear polarization Stokes parameters $Q$ and $U$ are given by: 

\begin{eqnarray} \label{eq_qu}
q = \epsilon \rho \cos 2\psi \sin^2 i, \nonumber \\
u = \epsilon \rho \sin 2\psi \sin^2 i,
\end{eqnarray}

\noindent
where $\rho$ is the local density and $i$ is the inclination of the local magnetic field with respect to the line of sight. We then obtain the integrated $Q$ and $U$, as well as the column density, along the LOS. The polarization degree is given by $p = \sqrt{Q^2+U^2}/I$ and the polarization angle by $\phi = \arctan(U/Q)$. Once projected into a given line of sight, the polarization map is related to the synthetic emission map. 

The numerical simulations do not include the effects produced by gravity, but we expect they are very well suited for characterizing the turbulent regimes into the envelopes. On the other hand, high density clumps, as well as other large emission pixels related to superimposed clouds along the line of sight, may contaminate the statistics of models and more particularly observational data. Therefore, after scaling of the dynamical range of the simulations, a variation study of the space parameter of the observed maps is investigated to remove high density regions of the filaments where cores are generally forming. 

\begin{figure}
\epsscale{1.}
\plotone{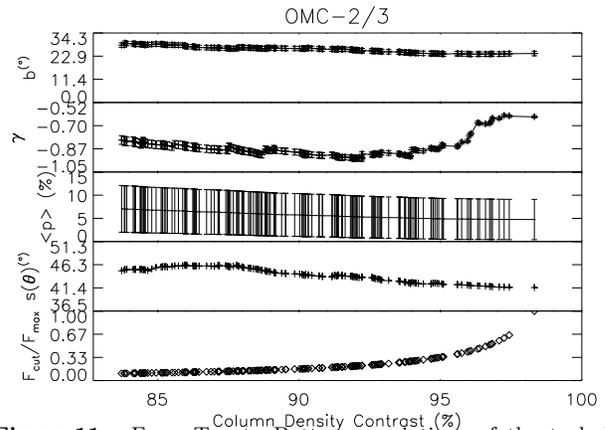}
\caption{From Top to Bottom: variations of the turbulent angular
  dispersion parameter $b$, of the depolarization parameter $\gamma$, 
of the mean polarization degree $<p>$, of the polarization angle
dispersion s($\theta_{p}$), and of the normalized flux treshold
F$_{\rm cut}/$F$_{\rm max}$ as a function of the Column Density
Contrast parameter for the OMC-2/3 region.
\label{var_omc23}}
\end{figure}

In order to avoid a bias related to rare very large emission pixels we studied the variations of the statistical parameters of the polarization maps $b$, $\gamma$, $<p>$ and $s(\theta_{p})$, in terms of the Column Density Contrast parameter, CDC = (Flux$_{\rm max}$ - Flux$_{\rm min}$)/Flux$_{\rm max}$. As an example, the variations of the four parameters obtained from the 
observed map of OMC-2/3 are plotted in 
Figure  \ref{var_omc23} as a function of CDC \citep[more details about the method are given by][]{poi10}.
Variations of parameter $s(\theta_{ p})$ are given for illustration purpose
only since, following the analysis given in section 2.4, parameter $b$ 
is the parameter suited for analyzing 
the effects of the turbulence at the resolution of the observations.

For estimating the values of the parameters that will be fitted by the simulations we need to identify 
the pixels that will be masked into the polarization maps. To do so we
first look to the variations of $\gamma$ with the CDC. For the OMC-2/3
regions $\gamma$ decreases drastically from CDC $\sim 96.5\%$ to CDC
$\sim 95.5\%$, but is quite constant for values of CDC $< 95.5 \%$. We
therefore adopt a limit value of the CDC of $\sim 95\%$ to define the
values of the parameters we will use for comparing the OMC-2/3 region
with simulations. The same strategy is followed for regions S106, W49
and DR21. 
The new values of $\gamma$, $b$ and $<p>$ extracted from the 
observed maps that will be compared with the values of the same parameters 
obtained from the analysis of simulated maps are displayed in Table \ref{ang_sum}. 

From the theoretical point of view, the statistics of polarization
maps is largely related to the turbulent regime of the cloud and to
the orientation of the mean magnetic field with respect to the line of
sight $\alpha$ (see \citet{fal08} for details). Therefore, once the
turbulent regime has been chosen (from the scaling discussed above), we
must study the dependence of the statistics of the polarization maps
with $\alpha$ and compare them to the observations. This procedure
allows us to constrain the orientation of the magnetic field from the
statistical parameters derived from the analysis of the polarization maps.

\begin{figure*}
\centering
\includegraphics[width=4cm]{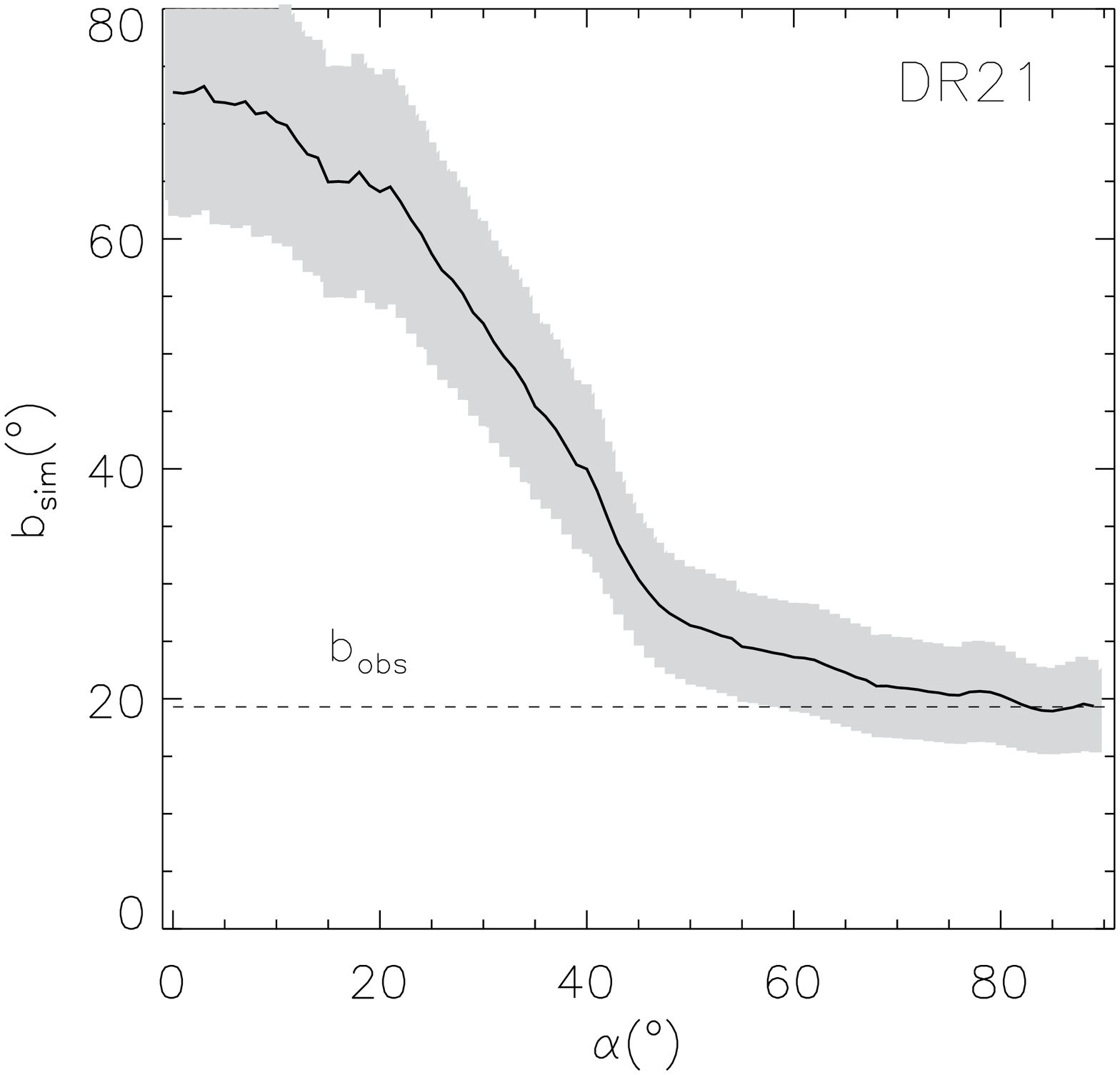}
\includegraphics[width=4cm]{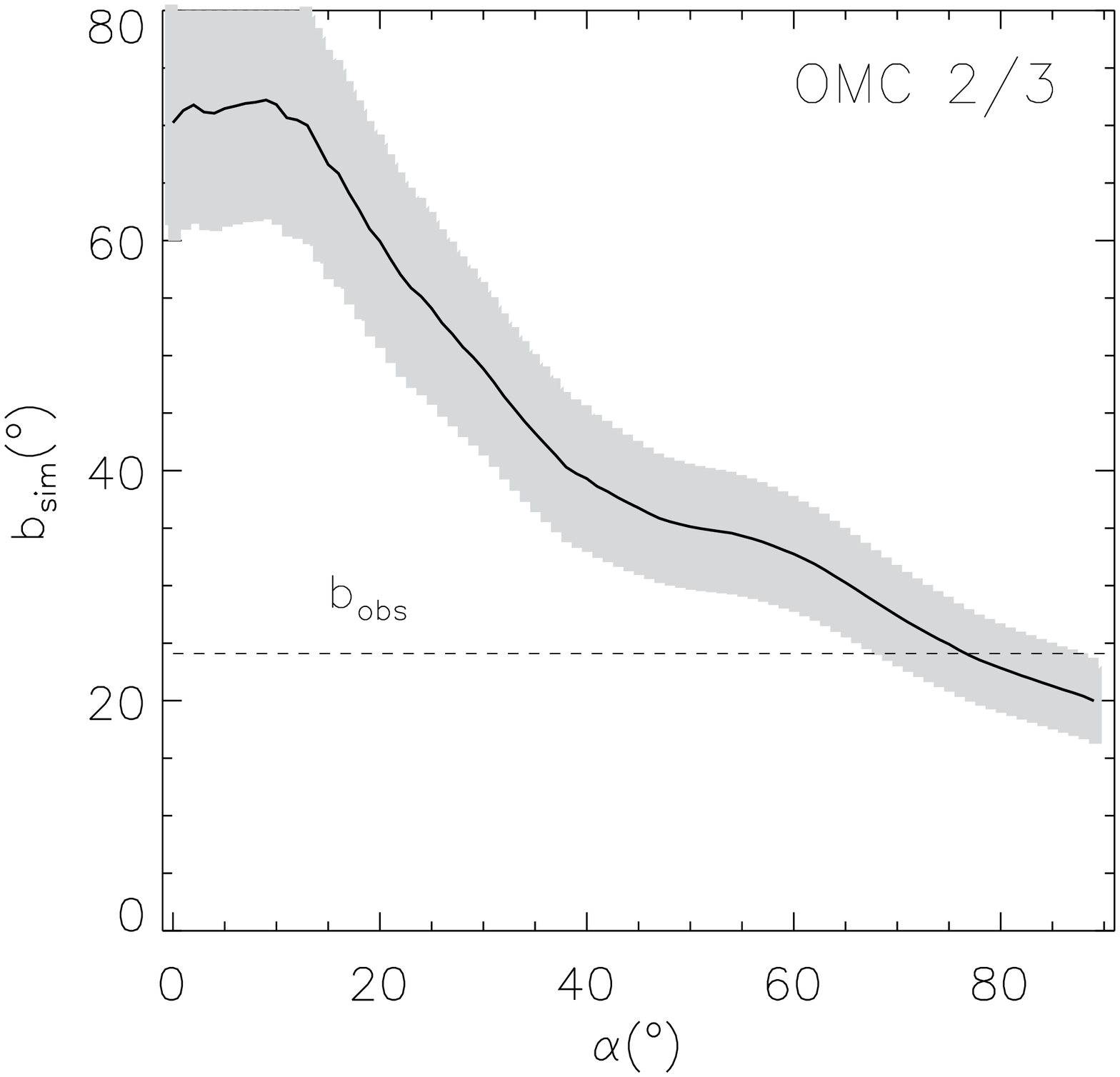}
\includegraphics[width=4cm]{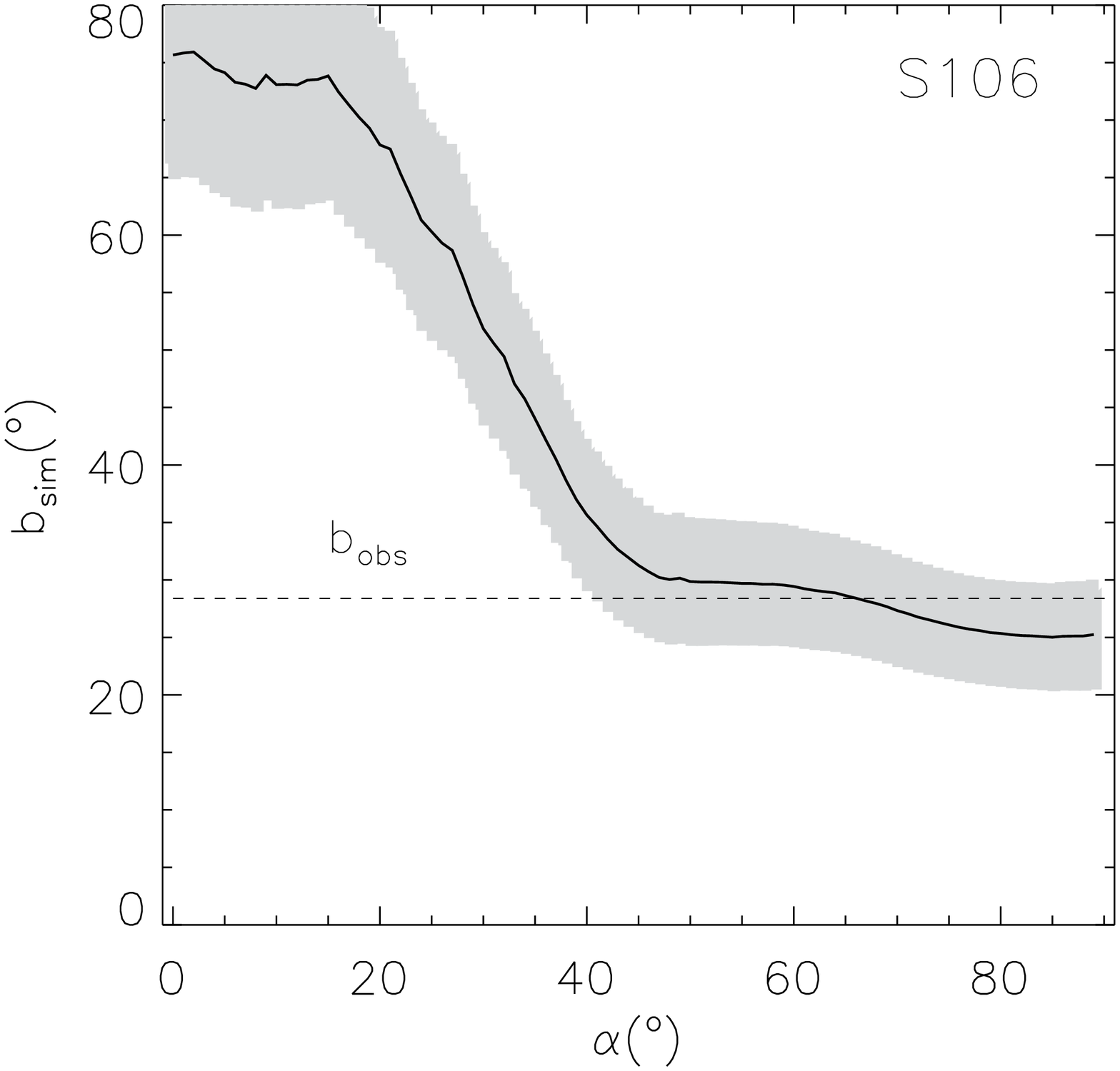}
\includegraphics[width=4cm]{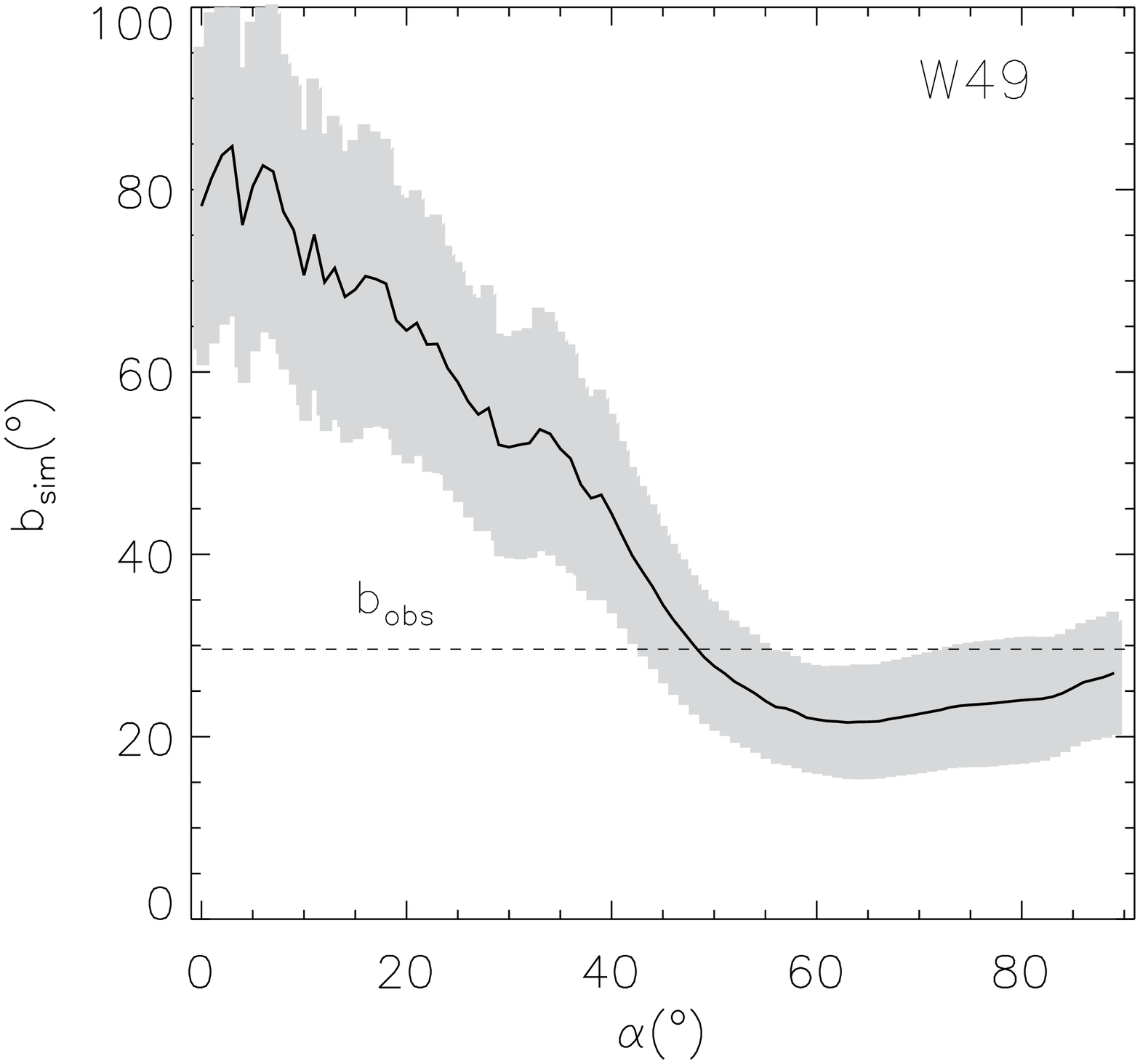}\\[10pt]
\includegraphics[width=4cm]{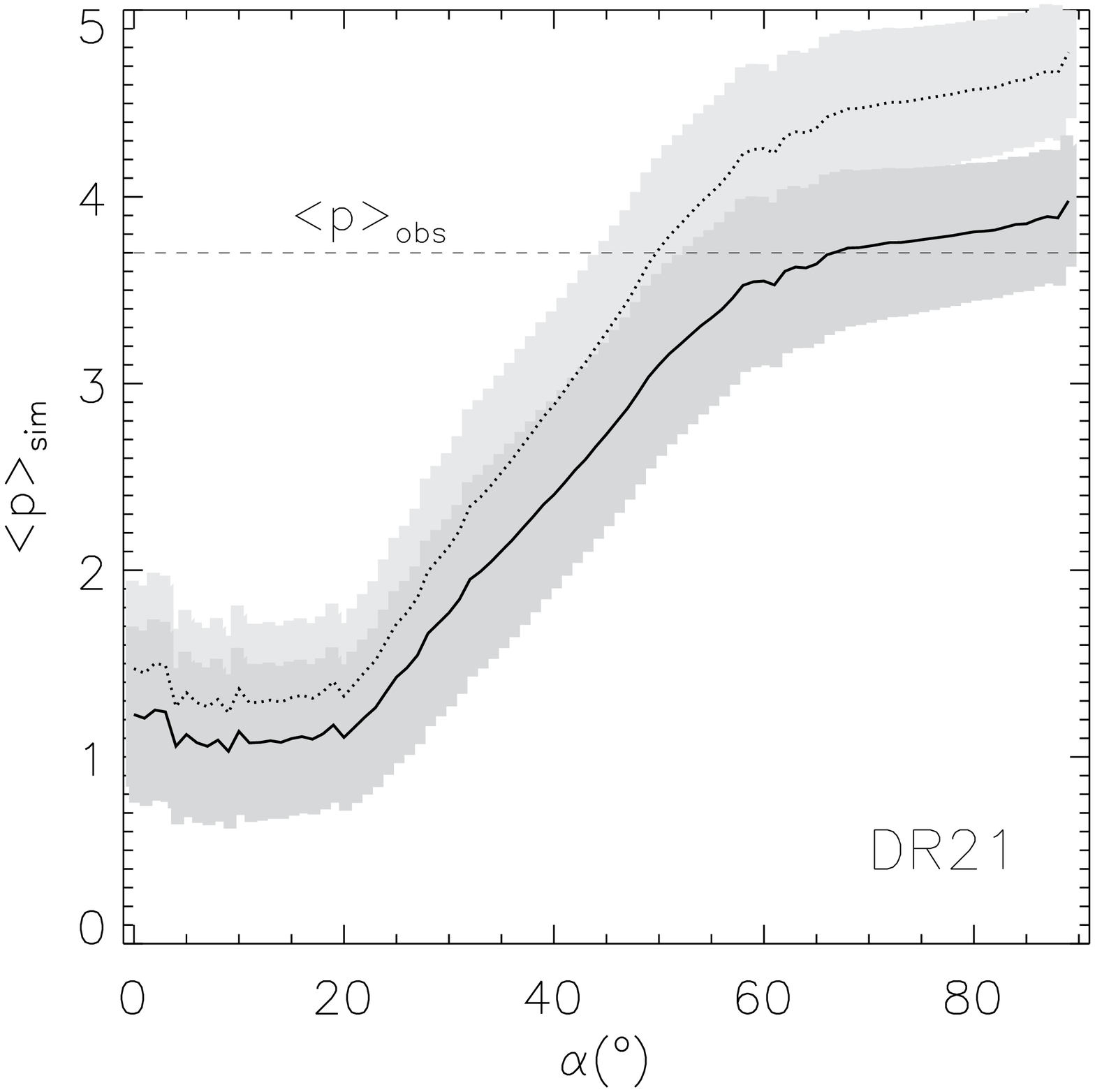}
\includegraphics[width=4cm]{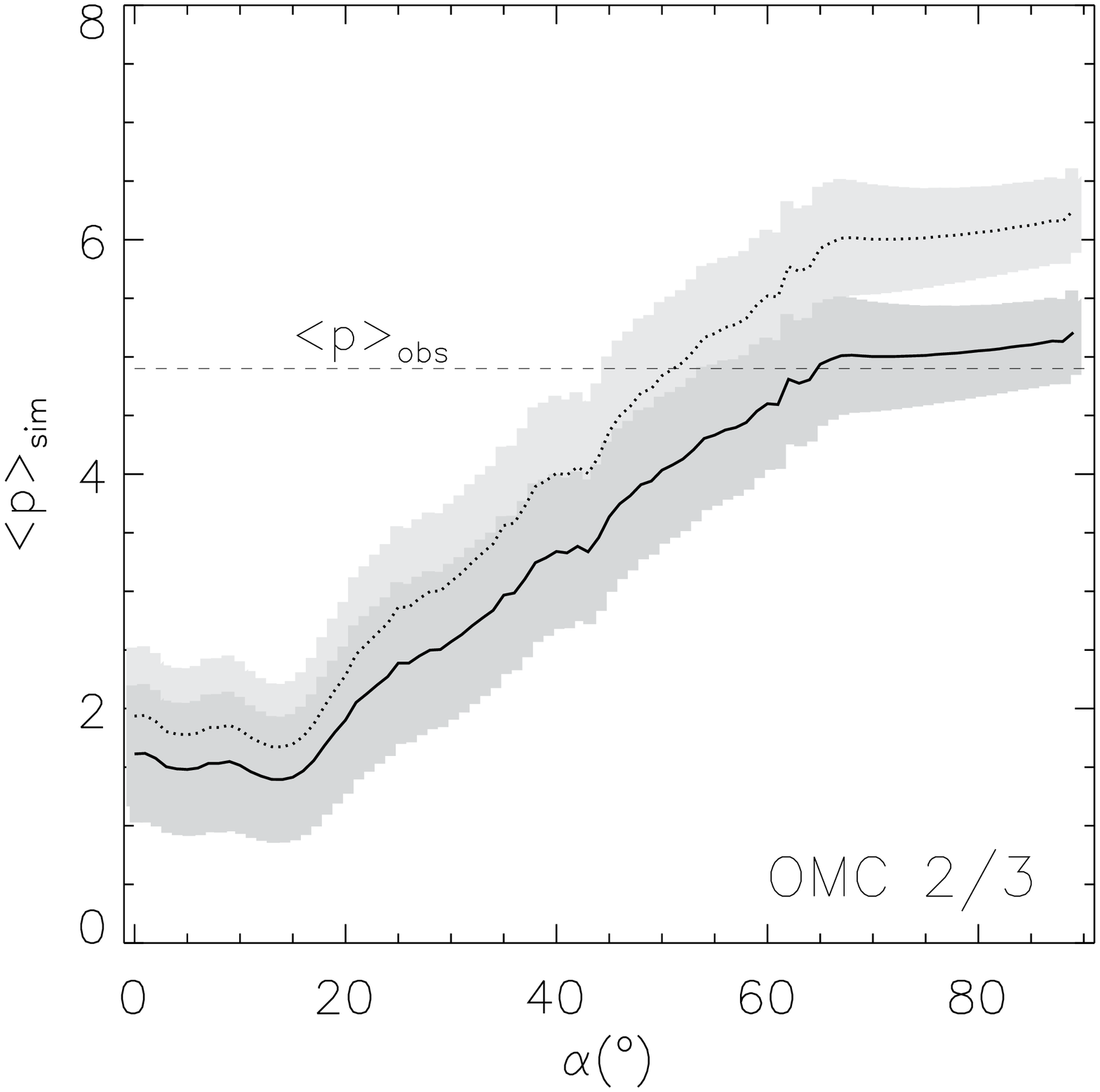}
\includegraphics[width=4cm]{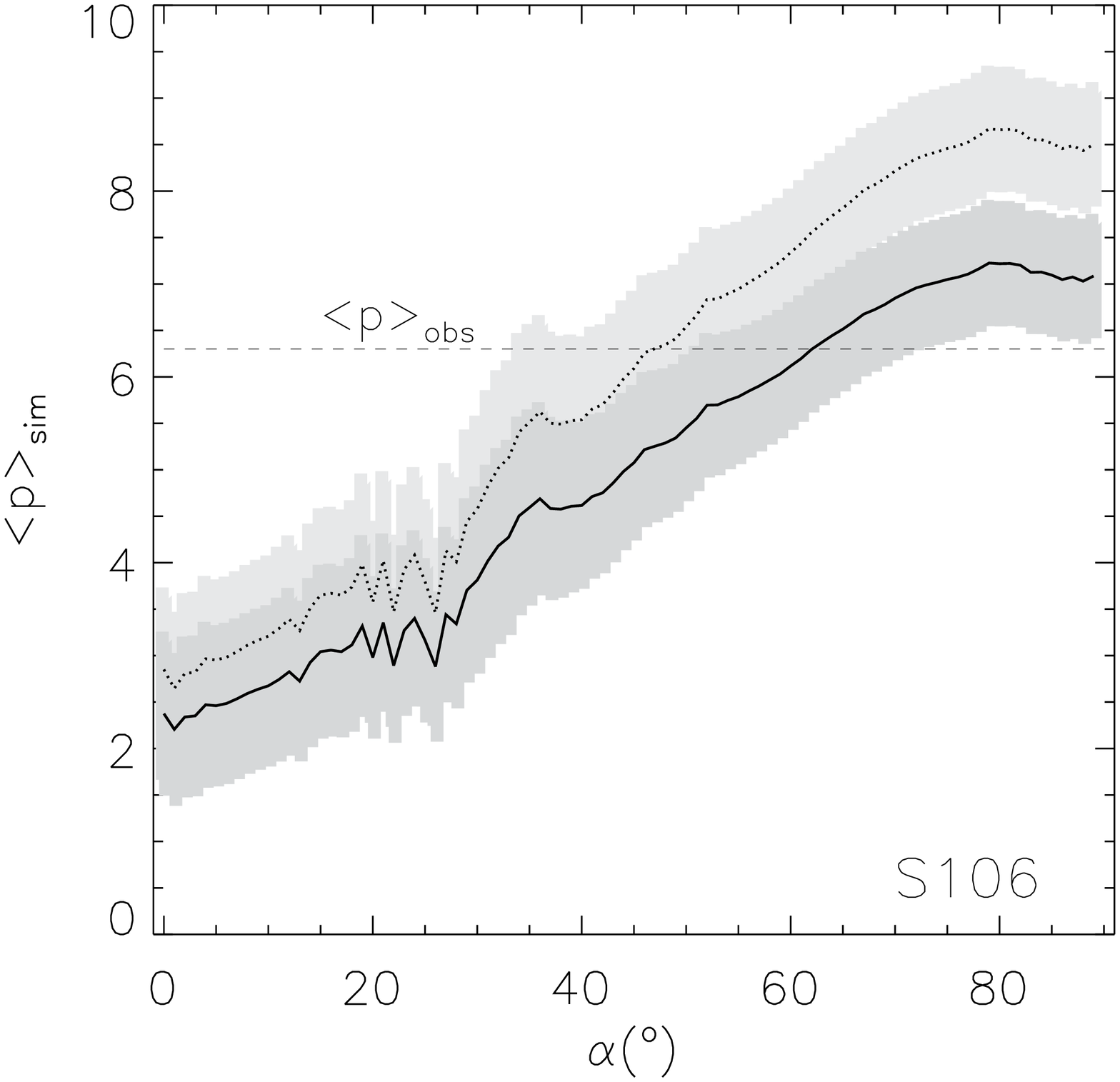}
\includegraphics[width=4cm]{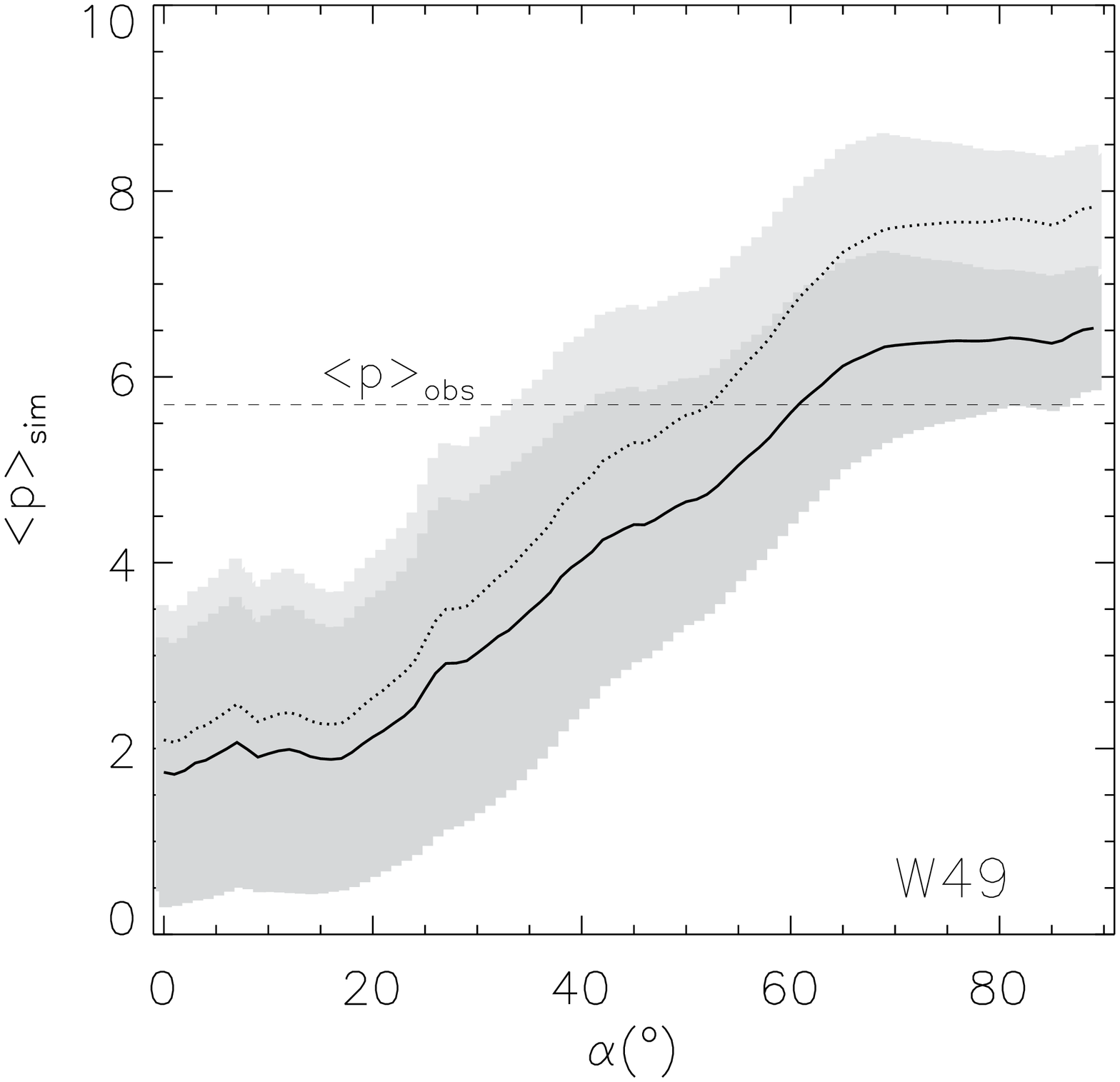}
\caption{Top plots: Variations of the  turbulent angular 
dispersion parameter $b_{\rm sim}$ as a function of the 
inclination angle $\alpha_{\rm sim}$ of the uniform initial magnetic field. 
The dotted line shows the value $b_{\rm obs}$ obtained from the
analysis of the maps of the different regions.
Bottom plots: same as top plots but for the mean polarization degree
parameter. The black full line shows results obtained with
$\epsilon =0.1$; the grey dashed line shows results 
obtained with $\epsilon =0.12$ (see text for details).
Intercepts of curves of $b_{\rm sim}$ and $<p_{\rm sim}>$ (including error bars) with
values of $b_{\rm obs}$ and $<p_{\rm obs}>$, respectively, give
constraints on values of $\alpha_{\rm sim}$.
}
\label{plotssim}
\end{figure*}

In Figure \ref{plotssim} we present $b_{\rm sim}$ and $<p_{\rm sim}>$ as extracted from the simulated maps in terms of $\alpha$. 
It is clear the relation of the average polarization and the decorrelation parameter $b_{\rm sim}$ with respect to the projection angle $\alpha$. For $\alpha \rightarrow 0$, most of the polarization arises from the random turbulent component of the magnetic field. Therefore, the depolarization effect due to the integration of the non-uniform component along the LOS is enhanced. Also, the decorrelation length for the structure function of the magnetic field vectors decreases and $b_{\rm sim}$ increases. As $\alpha \rightarrow 90^{\circ}$, the polarization degree increases, and as the decorrelation length increases $b_{\rm sim}$ decreases. We found a very similar trend for both $b_{\rm sim}$ and $\sigma_{\rm sim}(\theta)$ with respect to the orientation of the LOS. 

The values of the parameters $b_{\rm obs}$ and $<p_{\rm obs}>$
obtained from the analysis of 
the observations (see Table 5) are shown with horizontal 
dashed lines in Figure \ref{plotssim}.
There is no perfect match of $\alpha$ values obtained for each of the statistical parameters, as it may be related to biases induced by dense structures that still contribute for the
synthetic maps. With a constant polarization efficiency,
$\epsilon = 0.1$, for DR21, the observed values of $b_{\rm obs}$ and $<p_{\rm obs}>$ are obtained for
$\alpha_{\rm sim} > 60^{\circ}$. Similar comparisons done for W49, lead to $50^{\circ} < \alpha_{\rm sim} < 85^{\circ}$. 
For S106, we find $50^{\circ} < \alpha_{\rm sim} < 70^{\circ}$. For OMC2/3, we estimate $\alpha_{\rm sim} > 70^{\circ}$.

For every region, we find values of $\gamma > -0.5$, therefore it is
not possible to put any constraint on the angle $\alpha$ with 
this parameter. More discussion about this aspect of our work is given in the next section.


\section{DISCUSSION} \label{discussion}

\subsection{Shortcomings of the Simulations} \label{shortcomings}

Since we focused the modeling of polarization statistics on the
turbulent regime of the studied regions we deliberately avoided the
use of self-gravity and radiative cooling. For the molecular gas,
radiative cooling plays a minor role in the evolution of the turbulent
motions and the statistics of the dense structures, and result in very
little variations in local temperatures in the range of $T\sim 8 -
20$K. Self-gravity, on the other hand, plays a major role on the
smaller scale dynamics of the molecular clouds, resulting in their
fragmentation and collapse of Jeans unstable cores.  We tried to
remove possible effects of the cores by removing the largest column 
density pixels out of the statistical study and, as shown before, the 
parameters $\sigma(\theta)$, $<p>$ and $b$ are only slightly sensitive to 
that process. In this sense, except for parameter $\gamma$ which will
be discussed more in the following, our analysis does not rely on
parameters which are too sensitive to the structures of the cloud
considered. This can be seen in Figure \ref{var_omc23} where
variations of parameters $b$, $<p>$ and even $s(\theta)$ are
quite smooth as the CDC parameter decreases.

From the point of view of dust grains, our simulations do not take
into account possible variations of $p$ with parameters $\lambda$, 
T$_{g}$ and $\beta$, i.e. the wavelength of dust grain emission, the dust grain temperature and 
the dust grain emissivity index, respectively. Variations of dust
grains size and / or variations of dust grains axis ratio, whether
their shape is oblate or prolate on average, or non ideal, are also
not included explicitly. Nevertheless all this infotmation is statistically and
implicitly included in the choice of the value for parameter
$\epsilon$ used in equation \ref{eq_qu} to make our numerical
calculations. We will see in the next section that a value 
of $\epsilon \simeq 0.1$ is a reasonable statistical value to 
describe the turbulent states of the systems investigated.

\subsection{Polarization Efficiency Parameter} \label{poleff}

Two variants of the definition of the polarization efficiency
parameter (or polarization reduction factor) have been described and
used up to now. They have been mainly proposed by theoreticians based on
the fact that the measured linearly polarized signal, as supposed to be
produced by layers of aligned dust grains, whether in emission or by
dichroic extinction, seems never to exceed a given threshold. In the
submm range, which is the domain of wavelengths we are interested, 
this threshold is often considered to be around a value of 10 $\%$
based on ground-based or airborne experiment observations of
sufficiently bright star-forming regions \citep[e.g.][]{hil99}, while
maxima lying between 10 $\%$ to 20 $\%$ are expected in more diffuse
regions of the ISM \citep[see][]{ben04}.

The general definition of the polarization efficiency parameter is given by: 
\begin{equation} \label{eq_phi}
\Phi = R F \cos^{2} \gamma_{pos},
\end{equation} 
where  $F$ is the polarization reduction due to the turbulent
component of the magnetic field and $\gamma_{pos}$ is the angle between the POS and the 
local direction of the magnetic field while $R$ is the Rayleigh
polarization reduction factor. 

In the first variant of the definition provided by \citet{gre68}, $R$ is defined as:  
\begin{equation} \label{eq_rgreen}
R = 1.5(\langle \cos^{2} \beta \rangle -1/3)
\end{equation} 
where $\beta$ is the angle between the grain angular momentum vector and the magnetic field.
In this case $R$ translates as a constant imperfect alignment in the reference frame of each dust grain.

A second definition for $R$ has been proposed by \citet{cho05} such that:
\begin{equation} \label{eq_rlaz}
R=\frac{\int^{a_{\rm max}}_{a_{\rm alg}}C_{\rm ran}n(a)da}{\int^{a_{\rm max}}_{a_{\rm min}}C_{\rm ran}n(a)da},
\end{equation}
where, $C_{\rm ran} = (2 C_{\perp}+C_{\parallel})/3$ 
is the dust grain cross section, $a$ is the dust grain size, $n(a)$ is the grain number density, $a_{\rm min}$
is the minimum size, $a_{\rm max}$ is the maximum size, and $a_{\rm  alg}$ is the minimum aligned size.

This second definition differs from the first one in the sense that
$R$ is now defined as a function of the distribution in size of the
grains. Environment effects are very important and the central idea
is that non symetrical grains can be radiatively aligned by torques
\citep[Radiative Alignment by Torques or RATs, see][]{hoa08} 
produced by energy
transfer on dust grains during their interaction with the Interstellar
Radiation Field (ISRF). In particular big grains could be perfectly
aligned by this process.

One can now compare those definitions with the assumptions used in
and the calculations done with our simulations. 
Comparison of equations \ref{eq_qu} and \ref{eq_phi} 
shows that  $\gamma_{pos} = 90^{\circ} - i$, 
i.e. all simulated dust grains located in a cell of density $\rho$ are
identically aligned by a local magnetic field which orientation makes
an angle $\gamma_{pos}$ with the POS. The effects of the turbulence 
as described by parameter F in equation \ref{eq_phi} are directly estimated
with our simulations, 
and the Rayleigh parameter $R$ whether it is described by equation
\ref{eq_rgreen} or \ref{eq_rlaz}, can therefore be identified to our
parameter $\epsilon$ shown in equation \ref{eq_qu}.

In this context our choice of value for $\epsilon$, identified as
the maximum polarization degree that could be measured at a given
wavelength, should be interpreted as the case where optimal ideal
physical conditions are encountered in the ISM. Or in other words, 
the maximum polarization degree observed at 850 $\mu$m in the SCUPOL 
catalog would be the maximum value measured if the 
large scale magnetic field were uniform, lying in a plane almost
parallel to the POS and largely dominating over the turbulent field. 
In such a case we would, therefore, have $\phi  \approx \epsilon$.

Following this line of thought we looked into the values of the
maximum degree of polarization $p_{\rm max} $ measured on the 4 sources
compared with our simulations. All measurements are excessively high
with $p_{\rm max} > 19 \%$ for pixels where the flux density $I$ is each
time very low and all corresponding pixels are located on the edges
of the clouds. In fact, these estimates of $p_{\rm max} $ look like outliers in 
the polarization degree distribution which we 
believe can simply be explained by a high sensitivity of the values of
$p = \sqrt{Q^2+U^2}/I$ in these particular domains of the
maps due to the calibration of the intensity maps; i.e. $p$ would be 
very sensitive to small offsets on $I$ on the edge of the maps. 
Therefore, we have adopted another approach and 
checked how the value of  $\epsilon= 0.1$ compares with 
the statical maximum $p_{\rm max,stat} \sim < p >+ 3 \times s(p)$
which one 
could expect for the distributions of $p$ of each sample with 
the assumption that the distribution of the Stokes parameter is at first order  
gaussian in nature \citep[e.g.][]{ser62}.  
We find values of
$p_{\rm max,stat}$ equal to $\approx 7.8 \%$, $\approx 9.2 \%$, $\approx 8.5
\%$ and $\approx 5.4 \%$ 
for regions OMC-2/3, W49, S106 and DR21, respectively, i.e. values
which are
lower than 10 $\%$. These results do not justify our choice
for $\epsilon$, but show that this choice is not inconsistent with the
distribution of the polarization degree detected in each region. 
In addition, the mean turbulent-to-uniform ratios obtained from the 
SF analyses for those regions all show that the uniform components 
of the magnetic fields should dominate over the turbulent ones within
a factor of 2 to 3, i.e. the magnetic field turbulent component is not 
expected to constrain $<p>$ to values close to 0, in agreement 
with the results displayed in Table \ref{table_regions}.

On the other hand, a closer look to the SCUPOL catalog and to the
results displayed in Table 1 show that values of the mean polarization
degree $<p>$ can be found as high as $\approx 10 \%$, as in the case
for IRAM 04191+1522 and Mon IRAS 12. Therefore, one could expect 
variations of the polarization efficiency from one region to the
other. A look to the plots shown in Figure \ref{plotssim} show that 
variations of $\epsilon$ should only affect the variations of $<p_{\rm sim}>$
as a function of $\alpha_{\rm sim}$. For illustration purpose, we have 
added the curves shown with grey dashed-lines obtained for $\epsilon = 0.12$
and for each region, except OMC-2/3, 
we still find common fitting solutions to $\alpha$ 
from the intercepts of $b_{\rm sim}$ and $<p_{\rm sim}>$ 
with $b_{\rm obs}$ and $<p_{\rm obs}>$, respectively. 

In conclusion, if the physics employed in our simulations to describe the turbulent
regimes of the cloud envelopes is correct, as suggested by the results 
displayed in Table \ref{table_models_0p30}, then solutions to
$\alpha$ should be firstly constrained from the 
comparison of $b_{\rm sim}$ with $b_{\rm obs}$. And, on a second 
round, it would be theoretically possible to constrain 
a range of solutions to $\epsilon$ by exploring where the intercepts   
of $<p_{\rm sim}>$ with $<p_{\rm obs}>$ obtain for different values 
of $\epsilon$ give values of $\alpha$ still consistent with the solutions 
obtained from  the comparison of $b_{\rm sim}$ with $b_{\rm obs}$.

\subsection{Comparison with other Works}

\citet{cho05} show that under peculiar conditions depolarization could
occur if grains embedded in dark clouds are aligned by radiative
torques (RATs) such that their long axis is perpendicular to the
magnetic field. In their model, the mean field is about 2 times
stronger than the fluctuating magnetic field, a condition consistent
with our results obtained from the analysis of the four regions
compared with our simulations. The authors assume a uniform component 
of the magnetic field to be in the POS, i.e. $\alpha \approx 90^{\circ}$. 
Their results should be valid for clouds without embedded massive stars.

An application of the definition proposed by \citet{cho05} 
(see equation \ref{eq_rlaz}) has been investigated by Pelkonen et
al. (2007) with magnetohydrodynamic (MHD) simulations of turbulent
supersonic flows distinguishing sub-Alfv\'enic from super-Alfv\'enic
cases. The inclusion of a proper radiative transfer calculation
without detailed simulations of anisotropy shows that alignment
efficiency decreases as the RATs become less important into the denser
regions. In their work, the question is raised how to distinguish the
possible effect of the magnetic field topology from the interception
of multiple sources. A possible complication for which likelihood
should be avoided is by masking the high density regions 
as we did in our work. The maximum
degree of polarization used by \citet{pel07} is fixed to 15$\%$.

An extension of the \citet{cho05} model has been
proposed by \citet{bet07} by including the
effects of the mean Interstellar Radiation Field (ISRF), as well as
solutions of the radiative transfer equation to clumpy, optically
thick ($A_{\rm V} \sim 10$) prestellar cores and turbulent molecular
clouds. Their results are consistent with those described by
\citet{cho05}  and isothermality of large aligned grains is shown to
be a reasonable hypothesis. With the Rayleigh reduction factor 
$R$ defined from equation \ref{eq_rlaz}, those authors find that the maximum 
polarization degree, as well as the power index of the $p - I$ relations 
are extremely sensitive to the adopted upper cutoff of the power-law 
distribution of dust grain size, but only moderately sensitive to 
simplifying assumptions about the radiative anisotropy and no 
maximum polarization degree is arbitrarily introduced. Despite the highly 
complex topologies of the magnetic field, those authors show that 
polarization maps should trace the mass-weighted projected magnetic 
field vectors reasonably well. Their model explores a large area of
parameter space, but does not study the variations of MHD regimes 
nor consider the impact of embedded sources.

Supplementary analysis on the subject is given by \citet{pel09} where
an extension of their 2007 work is proposed. In their study they
consider $F=1$ and a maximum polarization degree still fixed 
to 15$\%$. The effect of the distribution of the size of the grains is 
investigated further away and the effects of the anisotropy of the
ISRF are added to the analysis. They find that the inclusion of
direction-dependent radiative torque efficiency weakens the dust 
grains alignment.
This effect can be partially counterbalanced if the grain size is doubled 
in denser regions which means that magnetic fields could theoretically 
still be probed up to $A_{\rm V} \sim 10$ in regions without embedded 
sources and where the 
dynamical timescale of coagulation processes is short enough.

The simulations discussed above cannot be compared directly to ours 
mainly because they do not explore the same MHD regimes as we do, 
and also because they are more focused on cloud cores physical
characterization rather than on cloud envelopes characterization, as we propose. 
For example, the turbulent regimes discussed by \citet{pel09} are 
supersonic and super-Alfv\'enic while the one we find
for the envelopes of S106, OMC-2/3, W49 and DR21 are supersonic and 
sub-Alfv\'enic. This might also explain the difference on the choice
of values for $\epsilon$.  

\subsection{The Depolarization Parameter}

The synthetic polarization maps obtained from the MHD simulations give 
estimates of $\gamma_{\rm sim} > -0.5$. These values are
systematically different
from the results obtained from the observed maps, where $\gamma_{\rm obs} < -0.5$. 
We can therefore speculate what could be the cause of this difference.

One of the main aspects that distinguish simulations and observations
is the instrument sensitivity. While in numerical simulations we
select the cells we want to use in the statistics, on the
observational side the degrees of freedom are much smaller. Low
intensity regions of the MHD simulations could be over-represented 
compared to real data sets because observed LOS low column density 
cannot be probed from the ground. As pointed in 
\citet{fal08}, at the lower end of intensity
range the simulations reveal no correlation between intensity and 
polarization degree. The reason for this is that the turbulence in the 
low intensity regions is basically sub-Alfvenic, i.e. magnetically 
dominated (see Burkhart et al. 2009). Once we move the statistics to 
the high intensity regions, the denser regions become more turbulent 
dominated, resulting in strong depolarization. 
{\it The turbulent depolarization is a function of the intensity}. 

From the observational point of view, the limited sensitivity of the
instrument biases the statistics for the high intensity regions. It
would be theoretically possible to remove low intensity pixels in the simulated
maps and produce steeper slopes of the power-law used to fit the $p-I$
relation but how to choose this threshold limit would be too much
arbitrary given the shortcomings of the models discussed in
section \ref{shortcomings}.

Another possible effect taking part is the alignment efficiency of the
dust particles. The numerical simulations do not include radiative
alignment for instance. \citet{bet07} and \citet{pel09} show that, with the definition of the Rayleigh reduction
factor $R$ from equation \ref{eq_rlaz}, the maximum polarization
degree as well as the power-law index of the $p-I$ relation are
extremely sensitive to the adopted cutoff of the power-law
distribution of dust grain size and moderately sensitive to
symplifying assumptions about the radiative anisotropy, as remarked before.  
At the same time, as mentionned before, their modelling do not explore
the same range of
turbulent regimes than we do which limits a direct comparison with our works.

\subsection{Interdependence of the Parameters}

Every probability distribution function of the 
polarization degree $p$, PDF($p$), can theoretically be 
described as a function of the parameter
$\Phi$ from equation \ref{eq_phi} with the following relation:
\begin{equation} \label{eq_interdep}
\rm PDF \it (p) = f (\Phi = R F \cos^{2} \gamma_{pos}).
\end{equation} 

For each regions, the estimate of $<p>$, is expected to depend
on the values of the Rayleigh polarization reduction factor $R$,
of the turbulent polarization reduction factor $F$, and 
on the average inclination of the uniform component of 
the magnetic field $\gamma_{pos}$.
Within our modeling the parameter $\gamma_{pos}$ is directly deduced 
from the initial conditions of the simulations.
The effect of $F$ is directly included in our
calculations via the integration of the Stokes parameters along each
LOS and estimates of $b$ should only depend on this parameter.
In addition, the $p-I$ relation power-law index 
parameter $\gamma$ should mainly depend on the values of the
parameter $R$. This seems to be consistent 
with the lack of correlation between parameters $p$, $b$ and $\gamma$
discussed in section \ref{parspace} and shown in Figures 
\ref{gamma_vs_b}, \ref{gamma_vs_pmean} and \ref{b_vs_stheta}.

On the other hand, if RATs is the dominating mechanism producing 
dust grain alignment, the interdependence of parameters $p$, $b$ and 
$\gamma$ might be more complex and for example several cases 
could be imagined such the that parameter $\gamma$ could 
depend on the values of $\Phi$ and not $R$ only.
Such studies are obviously out of the framework of this work 
and would need further investigations. 

\section{SUMMARY AND CONCLUSION} \label{summary}

In this work we have presented an extensive analysis of the star-forming and
molecular clouds 850 $\mu$m polarization maps 
of the SCUPOL Catalog produced by \citet{mat09}.

For each of the 27 sufficiently sampled regions, sets of parameters 
$<p>$, $b$ and $\gamma$ 
are systematically calculated in order to characterize the polarization properties, 
the depolarization properties and the turbulent-to-mean magnetic field 
ratio of each region as seen on the POS. 
As expected from theoretical modelling, the statistical
analysis showed no specific correlation between these parameters. 

We also created synthetic 2D polarization maps from 3D MHD 
$1024^3$ pixels grid simulations, performed for
different MHD regimes, as discussed by \citet{fal08}.  
Such MHD regimes are estimated for the S106, OMC-2/3, W49 and
DR21 molecular cloud regions with 3D MHD cubes
properly scaled to the observed maps. 
The values obtained from the simulations for the Alfv\'en and
sonic Mach numbers are in good agreement, within a factor of 2, with
the values obtained for those parameters from Zeeman measurements
as estimated by \citet{cru99} in the same regions. 

Constraints on the values of the inclination angle $\alpha$
of the mean magnetic field with 
respect to the LOS, are obtained by comparing the values of parameters 
$<p>$ and $b$ estimated from the simulated maps to those  
obtained from the observed maps of these four regions.
Last line of Table \ref{table_summary} gives a summary of the range of
estimates obtained for $\alpha$ from our data (1st line), 
and the constraints provided by \citet{hou04} (2nd line) 
and the range given by the combination of Zeeman measurements 
and the CF method (lines 3 and 4 of the Table).

Our main conclusion is that most of the results obtained from our
analysis of simple ideal isothermal and non-selfgraviting 3D MHD 
simulations, once properly scaled to the observations,  
are consistent with results obtained from the latter. This suggests
that turbulence only is sufficient to describe the 
basic dynamical properties of the molecular cloud
envelopes, without including the effects of gravity or 
radiation effects (e.g. in terms of grain alignment).

FOR ACKNOWLEDGEMENTS

We thank the anonymous referee for his comments that helped to improve
this work.
The research of FP has been supported by the Funda\c{c}\~ao de Amparo \`a Pesquisa do Estado de S\~ao Paulo 
(FAPESP no 2007/56302-4). FP also thanks the Leverhulme Trust through the Research Project Grant F/00 407/BN.
EDGP thanks the Brazilian agencies FAPESP (no. 2006/50654-3) and CNPq (306598/2009-4) for financial support.
DFG thanks the European Research Council (ADG-2011 ECOGAL), and 
brazilian agencies CNPq (no. 300382/2008-1), CAPES (3400-13-1) and FAPESP (no.2011/12909-8) for financial support.


\clearpage

\begin{deluxetable}{lcccccccccc} 
\tablewidth{0pt}
\tabletypesize{\scriptsize}
\tablecaption{Results of the Analysis. \label{table_regions}}
\tablehead{
\colhead{Object} &  \colhead{Region$^{(a)}$} &
\colhead{Map} & \colhead{Region$^{(b)}$} &
\colhead{$<p>$ $\pm$ s($(p)$)} & \colhead{$<\theta_{p}>$ $\pm$ s($\theta_{p}$)} &
\colhead{$\gamma$} & \colhead{$b$} &
\colhead{$\frac{<B^{2}_{t_{||}}>^{1/2}}{B_{0}}$} \\
\colhead{Name} &  \colhead{Type} &
\colhead{Pixel} & \colhead{Distance} & 
\colhead{} & \colhead{} &
\colhead{} & \colhead{} &
\colhead{} \\
\colhead{} & \colhead{} &
\colhead{Number} & \colhead{(kpc)} & 
\colhead{($\%$)} &
\colhead{$(^{\circ})$} &
\colhead{} & \colhead{$(^{\circ})$} &
\colhead{} \\  
\colhead{(1)} & \colhead{(2)} &
\colhead{(3)} & \colhead{(4)} & 
\colhead{(5)} & \colhead{(6)} &
\colhead{(7)} & \colhead{(8)} &
\colhead{(9)}   
}\startdata
      1 Galactic Center & GC &        654 &       8000.00 &        2.8 $\pm$ 0.6  &        142.1$\pm$       43.7 &      -0.94 $\pm$     0.01 &        24.6$\pm$       1.5 &       0.32$\pm$     0.02\\
       5 AFGL333 & SFR &        233 &       1.950 &        8.2 $\pm$ 1.2 &        001.3$\pm$       38.1 &      -1.00 $\pm$     0.05 &        21.1 $\pm$       1.5 &       0.27$\pm$     0.02\\
       8 NGC1333 & SFR &        193 &       0.320 &        6.2 $\pm$ 0.9&        055.1$\pm$       47.1 &      -0.26 $\pm$     0.02 &        35.1 $\pm$       1.5 &       0.48$\pm$     0.02\\
       9 Barnard 1 & SFR &        101 &       0.250 &        6.2 $\pm$ 1.0 &        054.9$\pm$       41.2 &      -0.83 $\pm$     0.05 &        37.5 $\pm$       1.4 &       0.52$\pm$     0.02\\
      11 OMC-1 & SFR &        385 &       0.414 &        4.0 $\pm$ 0.6&        155.4$\pm$       38.7 &      0.10 $\pm$    0.01 &        14.8 $\pm$       1.4 &       0.19$\pm$     0.02\\
      12 OMC-2/3 & SFR &        361 &       0.414 &        4.8 $\pm$ 1.0 &        079.5$\pm$       41.5 &      -0.63$\pm$    0.01 &        24.2 $\pm$       1.1 &       0.31$\pm$     0.02\\
      13 NGC 2024 &  SFR &        203 &       0.400 &        4.1 $\pm$ 0.8 &        092.1$\pm$       38.5 &      -0.84$\pm$     0.03 &        14.4 $\pm$       1.7 &       0.18$\pm$     0.02\\
      15 NGC 2068 & SFR &        285 &       0.400 &        7.0 $\pm$ 0.9 &        039.3$\pm$       39.4 &      -0.75$\pm$     0.04 &        35.1 $\pm$       1.4 &       0.48$\pm$     0.02\\
      18 Mon R2 IRS1 & SFR &        183 &       0.950 &        4.9 $\pm$ 0.9&        072.4$\pm$       30.7 &       -1.16$\pm$     0.03 &        21.9$\pm$       1.7 &       0.28$\pm$     0.02\\
      20 Mon IRAS 12 & SFR &        171 &       0.800 &        9.9 $\pm$ 2.1 &        049.7 $\pm$       45.8 &      -0.86 $\pm$     0.03 &        44.5$\pm$       1.1 &       0.66$\pm$     0.02\\
      22 rho Oph A & SFR &        337 &       0.139 &        5.3 $\pm$ 0.9&        063.2$\pm$       39.8 &      -0.80$\pm$     0.02 &        21.8 $\pm$       1.5 &       0.28$\pm$     0.02\\
      24 rho Oph B2 &  SFR &        113 &       0.139 &        5.3 $\pm$ 1.1 &        056.6$\pm$       45.1 &      -0.86$\pm$     0.06 &        36.2$\pm$       1.3 &       0.50 $\pm$     0.02\\
      25 NGC 6334A & SFR &        77 &       1.700 &        4.2 $\pm$ 1.0 &        094.8$\pm$       43.2 &      -0.82$\pm$     0.05 &        14.5 $\pm$       1.7 &       0.18$\pm$     0.02\\
      26 G011.11-0.12 & SFR  &        143 &       3.600 &        5.4 $\pm$ 1.2 &        134.7$\pm$       26.7 &      -0.89$\pm$     0.03 &        25.3$\pm$       1.6 &       0.33$\pm$     0.02\\
      27 GGD 27 &  SFR &        49 &       1.700 &        6.7 $\pm$ 1.6 &        178.1$\pm$       39.9 &      -0.82$\pm$     0.06 &        37.3 $\pm$       1.5 &       0.52$\pm$     0.03\\
      29 Serpens Main Core & SFR &        231 &       0.310 &        7.1 $\pm$ 1.4 &        075.3$\pm$       40.4 &       -1.01$\pm$     0.03 &        36.8$\pm$       1.3 &       0.51$\pm$     0.02\\
      33 W48 & SFR &        122 &       3.400 &        4.9 $\pm$ 1.0&        152.3$\pm$       44.2 &      -0.59 $\pm$     0.02 &        35.7 $\pm$       1.3 &       0.49$\pm$     0.02\\
      35 W49 & SFR &        368 &       11.400 &        5.3 $\pm$ 1.3&        108.5$\pm$       45.5 &      -0.83$\pm$    0.01 &        29.6 $\pm$       1.2 &       0.39$\pm$     0.02\\
      36 W51 & SFR &        123 &       7.500 &        3.2 $\pm$ 0.8 &        137.5$\pm$       50.5 &      -0.71$\pm$     0.03 &        34.0$\pm$       1.4 &       0.46$\pm$     0.02\\
      41 S106 &  SFR &        201 &       0.600 &        5.8 $\pm$ 0.9 &        115.7$\pm$       47.1 &      -0.78$\pm$     0.03 &        28.4$\pm$       1.4 &       0.37$\pm$     0.02\\
      43 DR21 & SFR &        439 &       3.000 &        3.6 $\pm$ 0.6 &        077.0$\pm$       35.9 &      -0.50$\pm$     0.01 &        19.3 $\pm$       1.8 &       0.25$\pm$     0.02\\
      47 S152 & SFR &        123 &       5.000 &        5.6$\pm$ 1.0 &        081.8$\pm$       34.3 &      -0.61$\pm$     0.04 &        21.2 $\pm$       1.8 &       0.27$\pm$     0.02\\
      50 L1448 & YSO &        130 &       0.250 &        5.1 $\pm$ 1.0 &        040.3$\pm$       47.5 &      -0.87$\pm$     0.04 &        41.1 $\pm$       1.4 &       0.59$\pm$     0.03\\
      53 L1527 & YSO &        92 &       0.140 &        6.6 $\pm$  1.3 &        101.4 $\pm$       38.7 &       -1.10$\pm$     0.05 &        37.4 $\pm$       1.3 &       0.52$\pm$     0.02\\
      54 IRAM 04191+1522 & YSO &        55 &       0.140 &        10.6 $\pm$ 2.5 &        074.5 $\pm$       35.3 &       -1.35$\pm$     0.09 &        35.2$\pm$       1.3 &       0.48$\pm$     0.02\\
      59 Cep A & YSO &        100 &       0.730 &        4.6 $\pm$  1.0 &        127.6 $\pm$       44.1 &      -0.65$\pm$     0.03 &        15.8 $\pm$       1.5 &       0.20$\pm$     0.02\\
      66 L43 & SPC &        40 &      0.170 &        7.2 $\pm$ 1.6 &        134.9$\pm$       37.3 &       -1.44$\pm$      0.14 &        33.1 $\pm$       1.4 &       0.45$\pm$     0.02\\
\enddata
\tablenotetext{(a)}{GC = Galactic Center; SFR = Star Forming Region; YSO = Young Stellar Object; SPC = Starless Prestellar Core.} 
\tablenotetext{(b)}{Distances are from \citet{mat09} and references therein.} 
\end{deluxetable}

\clearpage

\begin{deluxetable}{cccccc} 
\tablewidth{0pt}
\tabletypesize{\scriptsize}
\tablecaption{Sample Statistics. \label{table_statistics_regions}}
\tablehead{
\colhead{Sample} &  \colhead{Objects} &
\colhead{$\overline{<p>}$} & \colhead{$\overline{\gamma}$} & 
\colhead{$\overline{s(\theta_{p})}$} & \colhead{$\overline{b}$} \\
\colhead{} & \colhead{Number} &
\colhead{$(\%)$} & \colhead{} & 
\colhead{($^{\circ}$)} & \colhead{($^{\circ}$)}  
}\startdata
All Selected Regions& 27 &  5.72 $\pm$ 1.84 & -0.81 $\pm$ 0.30 & 40.7 $\pm$  5.4&       28.7$\pm$  8.9\\
Galactic Center     & 1  &  2.77 $\pm$ 0.00 & -0.94 $\pm$ 0.00 & 43.7 $\pm$  0.0&       24.6$\pm$  0.0\\
SFRs$^{(a)}$        & 21 &  5.59 $\pm$ 1.59 & -0.74 $\pm$ 0.27 & 40.6 $\pm$  5.7&       28.0$\pm$  8.9\\
YSOs$^{(a)}$        & 4  &  6.73 $\pm$ 2.72 & -0.99 $\pm$ 0.30 & 41.4 $\pm$  5.4&       32.4$\pm$  11.3\\
SPCs$^{(a)}$        & 1  &  7.23 $\pm$ 0.00 & -1.44 $\pm$ 0.00 & 37.3 $\pm$  0.0&       33.1$\pm$  0.0\\
\enddata
\tablenotetext{(a)}{SFR = Star Forming Region; YSO = Young Stellar Object; SPC = Starless Prestellar Core.} 
\end{deluxetable}
 

\begin{deluxetable}{cccccccc} 
\tablewidth{0pt}
\tabletypesize{\scriptsize}
\tablecaption{Total Magnetic Fields Intensities and Inclinations. \label{table_bfields}}
\tablehead{
\colhead{Object} & \colhead{$B_{\rm LOS}^{(a)}$} &  \colhead{log $n^{(a)}$} &
\colhead{$\sigma_{v}^{(a)}$} & \colhead{$B_{\rm POS}^{HIL09}$} & \colhead{$B_{\rm POS}^{FLK08}$} & \colhead{Map Areas Ratio} &
\colhead{$\alpha_{(\rm max)}^{(b)}$} 
\\
\colhead{Name} & \colhead{($\mu$G)} &
\colhead{(H$_{2}$ cm$^{-3}$)} & \colhead{(km/s)} & \colhead{($\mu$G)} & \colhead{($\mu$G)} & \colhead{(pol./spectro.)} 
& \colhead{($^{\circ}$)} 
}\startdata
OMC-1 &360& 5.9 (5.9)$^{(c)}$  & 0.60 (1.85)$^{(c)}$ & 1976 (6149)$^{(d)}$ & 395 (1229)$^{(d)}$ & $\sim$ 33&  79.7 (86.6)$^{(d)}$\\ 
NGC 2024  &87 & 5.0 (5.0)$^{(c)}$  &    0.64 (0.68)$^{(c)}$   & 773 (830)$^{(d)}$ & 154 (165)$^{(d)}$ & $\sim$ 2 & 83.6 (84.0)$^{(d)}$\\ 
$\rho$ Oph. A (1$^{(a)}$)  &10 & 3.0 (5.0)$^{(c)}$  &  0.55 (0.58)$^{(c)}$  & 43 (455)$^{(d)}$ & 8 (84)$^{(d)}$& $\sim$ 1 & 77.0 (88.7)$^{(d)}$\\ 
$\rho$ Oph. B2 (2$^{(a)}$) &14 & 3.2   &  0.59   & 33 & 2  & $\sim$ 1 & 66.9 \\ 
W49  &21 & 3.0   &    0.64    & 36 & $1-2$ & $\sim$ 0.1 & 59.4 \\ 
S106 &400& 5.3   &    0.68   & 562 & 19 & $\sim$ 16 & 54.6\\ 
DR21 (OH1$^{(a)}$) &710& 6.3 (6.3)$^{(c)}$ &  0.98 (2.04)$^{(c)}$  & 3897 (8140)$^{(d)}$ & 1210 (2528)$^{(d)}$ & $\sim$ 26 & 79.7 (85.0)$^{(d)}$\\ 
DR21 (OH2$^{(a)}$) &360& 6.0 (6.0)$^{(c)}$ &   0.98 (2.04)$^{(c)}$  & 2759 (5762)$^{(d)}$ &  856 (1789)$^{(d)}$ & $\sim$ 26 & 82.6 (86.4)$^{(d)}$\\ 
\enddata
\tablenotetext{(a)}{Values with no parenthesis are from \citet{cru99}
  based on CO or OH spectroscopy data.} 
\tablenotetext{(b)}{Upper limit estimate of the inclination angle of the mean magnetic field with respect to the LOS.} 
\tablenotetext{(c)}{With parenthesis: values used by considering
  H$^{13}$CO$^{+}$(J=3-2) as a tracer for $n$ and $\sigma_{v}^{(a)}$,
  see text in section \ref{zeeman} for details. Values $n=10^{-5}$ are
used if $n($CO or OH$)<10^{-5}$, otherwise $n($CO or OH$)$ estimates are used. } 
\tablenotetext{(d)}{No parenthesis: values obtained by considering CO
  or OH
  as a tracer for $n$ and $\sigma_{v}$. With parenthesis: results obtained by considering
  H$^{13}$CO$^{+}$(J=3-2) as a tracer for $n$ and $\sigma_{v}$.} 
\end{deluxetable}
 

\begin{deluxetable}{cccccc} 
\tablewidth{0pt}
\tabletypesize{\scriptsize}
\tablecaption{Description of the Simulations - MHD, 1024$^{3}$. \label{table_models_0p30}}
\tablehead{
\colhead{Model} & \colhead{$M_{S}^{(a)}$} & \colhead{$M_{A}^{(b)}$} & \colhead{Object Name} & \colhead{$M_{S}^{(c)}$} & \colhead{$M_{A}^{(c)}$}
}
\startdata
1..........&  2.0    &  0.3     &  S106 &  3.6 &  0.2 \\
2..........&  3.0    &  0.5     &  OMC-2/3 &  - &  - \\
3..........&  5.0    &  0.7     &  W49 &  5.9 &  0.6 \\
4..........&   5.0    &  0.7     &  DR21 &  4.0 (4.0) &  1.3 (1.8)$^{(d)}$ \\
\enddata
\tablenotetext{(a)}{Sonic Mach Number ($M_{S}$).} 
\tablenotetext{(b)}{Alfv\'enic Mach Number ($M_{A}$).} 
\tablenotetext{(c)}{values of Alfv\'enic and sonic Mach numbers given by \citet{cru99}.}
\tablenotetext{(d)}{values are for DR21 OH1 and OH2, the last being placed in the parenthesis} 
\end{deluxetable}

\clearpage

\begin{deluxetable}{ccccccccc} 
\tablewidth{0pt}
\tabletypesize{\scriptsize}
\tablecaption{Parameter extracted from the analysis of the observed maps for comparing with simulated maps. \label{ang_sum}}
\tablehead{
\colhead{Region} &\colhead{Cloud size}  & \colhead{Obs. resolution} & \colhead{CDC cut$^{(a)}$} & \colhead{Fraction of rejected}   & \colhead{$b$} & \colhead{$\gamma$} & \colhead{$P$} & \colhead{s$(\theta_{p})$}  
\\
\colhead{Name} & \colhead{(pc$\times$pc)} & \colhead{(pc)}  & \colhead{$(\%)$} & \colhead{pixels} &   \colhead{$(^{\circ})$} & \colhead{} & \colhead{$(\%)$} & \colhead{$(^{\circ})$}
}
\startdata
S106         & $\sim$ 0.9 $\times$ 0.6  & $\sim$ 0.03 &  85.3 & 25/201    &  $\sim$ 28.4 $\pm$ 1.4 & -0.91 $\pm$ 0.05 & 6.3 $\pm$ 3.4 & 47.3 \\
OMC-2/3      & $\sim$ 2.2 $\times$ 0.4  & $\sim$ 0.04 &  94.9 & 19/361    &  24.1 $\pm$ 1.1        & -0.86 $\pm$ 0.02 & 4.9 $\pm$ 4.4 & 42.3 \\
W49          & $\sim$ 16.6 $\times$ 9.9 & $\sim$ 0.55 &  95.4 & 36/368    &  $\sim$ 29.6 $\pm$ 1.2 & -1.05 $\pm$ 0.01 & 5.7 $\pm$ 5.7 & 44.7 \\
DR21         & $\sim$ 4.4 $\times$ 2.6  & $\sim$ 0.15 &  95.9 & 27/439    &  $\sim$ 19.3 $\pm$ 1.8 & -0.63 $\pm$ 0.02 & 3.7 $\pm$ 2.8 & 34.1 \\ \hline
\enddata
\tablenotetext{(a)}{See Figure 
\ref{var_omc23} for an example on region OMC-2/3 of the variation of the parameters as a function of the CDC. 
} 
\end{deluxetable}
 

\begin{deluxetable}{lcccc} 
\tablewidth{0pt}
\tabletypesize{\scriptsize}
\tablecaption{Mean Magnetic Field Inclination Angle Summary Table. \label{table_summary}}
\tablehead{
\colhead{Region Name} & \colhead{S106} & \colhead{OMC-2/3} &
\colhead{W49} & \colhead{DR21}
}
\startdata
$ \alpha_{\rm sim} (\epsilon=0.10)$  ($^{\circ}) $ (Envelopes) &  [50-70]    &$>$70     & [50-85]  & $>$60\\
$ \alpha_{\rm Ion/Neutral}$  ($^{\circ}) $  (Cores) &  ...  &  [72-80] $^{(a)}$   & ...  & [55-70]$^{(b)}$ \\
$ \alpha_{\rm CF (CO) + Zeeman}$  ($^{\circ}) $  (Cores + envelopes) & $<$55&  ...   & $<$ 60  & $<$ 83\\
$ \alpha_{\rm CF (H^{13}CO^{+}(J=3-2)) + Zeeman}$  ($^{\circ}) $  (Cores + envelopes) & ... &  ...   & ...  & $<$ 87\\
\hline
$ \alpha_{\rm All Methods} (\epsilon=0.10)$  ($^{\circ}) $  & [50-55] & [72-80]    & [50-60]  & [60 - 70]  \\
\enddata
\tablenotetext{(a)}{\citet{hou04}} 
\tablenotetext{(b)}{\citet{kir09}} 
\end{deluxetable}

\end{document}